\documentclass[aps, pre, twocolumn,showkeys,showpacs,amsmath,amssymb,superscriptaddress,notitlepage,floatfix]{revtex4-1}

\usepackage{lineno,hyperref}
\usepackage{graphicx,graphics}
\usepackage{dcolumn}
\usepackage{amsmath,amssymb,amsfonts}
\usepackage{latexsym,verbatim}
\usepackage{bm,bbold}
\usepackage{color}
\usepackage{ulem}
\usepackage[percent]{overpic}
\usepackage[font=small,labelfont=bf]{caption}
\usepackage{subcaption}

\newcommand{\avg}[1]{\langle #1 \rangle}

\newcommand{\loss}{\mathcal{L}}

\begin{document}
\title{Fluctuations in pedestrian dynamics routing choices}
\author{A. Gabbana}
\affiliation{Department of Applied Physics, Eindhoven University of Technology, 5600 MB Eindhoven, The Netherlands}
\author{F. Toschi}
\affiliation{Department of Applied Physics, Eindhoven University of Technology, 5600 MB Eindhoven, The Netherlands}
\affiliation{CNR-IAC, Via dei Taurini 19, 00185 Roma, Italy}
\author{P. Ross}
\affiliation{Studio Philip Ross, 5641 JA Eindhoven, The Netherlands}
\author{A. Haans}
\affiliation{Human Technology Interaction,  Eindhoven University of Technology, 5600 MB Eindhoven, The Netherlands}
\author{A. Corbetta}
\affiliation{Department of Applied Physics, Eindhoven University of Technology, 5600 MB Eindhoven, The Netherlands}
\begin{abstract}
  Routing choices of walking pedestrians in geometrically complex
  environments are regulated by the interplay of a multitude of
  factors such as local crowding, (estimated) time to destination,
  (perceived) comfort. As individual choices
  combine, macroscopic traffic flow patterns emerge. Understanding the
  physical mechanisms yielding macroscopic traffic distributions in
  environments with complex geometries is an outstanding scientific
  challenge, with implications in the design and management of crowded
  pedestrian facilities.
  In this work, we analyze, by means of extensive real-life pedestrian
  tracking data, unidirectional flow dynamics in an asymmetric
  setting, as a prototype for many common complex geometries. 
  Our environment is composed of a main walkway and a
  slightly longer detour. Our measurements have been collected during
  a dedicated high-accuracy pedestrian tracking campaign held
  in Eindhoven (The Netherlands).  
  We show that the dynamics can be quantitatively modeled by introducing 
  a collective discomfort function, and that fluctuations
  on the behavior of single individuals are crucial to correctly recover
  the global statistical behavior.
  Notably, the observed traffic split substantially
  departs from an optimal, transport-wise, partition, as the global
  pedestrian throughput is not maximized.
\end{abstract}
\maketitle

%=================================================================
\section{Introduction}\label{sec:intro}
%=================================================================

Countless daily-life scenarios entail pedestrians walking
towards a common destination and choosing among alternative
neighboring routes. Consciously or unconsciously, and in connection
with factors such as crowd density, estimated time to
destination, path directness~\cite{hughes2003flow}, 
perceived comfort/safety, background knowledge, habits or even aesthetics, 
each individual selects and walks a preferred route~\cite{seneviratne-tpt-1985,hoog2,
hoogendoorn-tr-2004,brown-eb-2007,
mehta-jou-2008, guo-jotg-2013, shatu-jotg-2019,sevtsuk-ijst-2020}. 

At the individual microscale level, 
the routing choice has been quantitatively modeled in terms of discomfort 
functional, $\mathcal{L}$, that individuals 
seek to minimize~\cite{hoogendoorn-tr-2004,campanella-trr-2009}. 
From a microscopic description it is possible to derive the 
macroscale behavior of a crowd, as in the 
model introduced by Hughes~\cite{hughes-trpb-2002},
where the connection between the Fermat principle (i.e. minimization 
of optical paths) and a macroscopic Eikonal description is
used, however neglecting individual variability.

In this paper, we show that random fluctuations 
at the single individual scale are key 
to recover the observed macroscale statistics.
We model the decision process
via a global (i.e. coupling all pedestrians)
variational minimization, showing how
crowd flows stem from the 
combination of the routing decisions operated concurrently 
by single individuals, comparing with 
data from a real-life pedestrian tracking campaign.

We consider a crowd of $N$ pedestrians, and define
a discomfort $\mathcal{L}$ depending on 
the (perceived) density $\rho$, time to destination $\tau$, and
path length $\lambda$ (and possibly other quantities), 
for each single individual.
In other words, $\mathcal{L}$ represents a functional 
defined on the crowd as a whole, entailing the state of each 
pedestrian. 

Understanding qualitatively and quantitatively the physical processes
that link (the statistics of) microscopic dynamics and the macroscopic
crowding patterns that these generate is an outstanding challenge. On
one side, this shares deep connections with active matter
physics~\cite{RevModPhys.85.1143}, where optics-like variational
principles
succeeded at describing dynamics of living agents (e.g.~ant
trails~\cite{oettler-plos-2013}). On the other side, physics-based
modeling of crowd dynamics retains great relevance in the
endeavor to increase safety and comfort of urban infrastructures and
large-scale events~\cite{leyden-ajph-2003,blanco-pp-2009}.

Among the factors undermining our understanding of crowd flows is the
inherent technical challenge of collecting accurate measurements at
large spatial and time scales.  Thus, the majority of the studies in
pedestrian dynamics have leveraged on qualitative
simulations~\cite{cristiani-amm-2019} via
microscopic~\cite{helbing1995PRE,helbing-nat-2000,blue-trr-1998,blue-trb-2001}
or macroscopic numerical
models~\cite{hughes-mcs-2000,treuille-tog-2006,duives-trc-2013}. 
Routing has also been addressed via questionnaires
(e.g.~\cite{borgers-ga-1986,verlander-tpm-1997,koh-jep-2013}) or in laboratory
conditions~\cite{kretz-josm-2006,kretz-josm-2006b,seyfried-ped-2010,
  moussaid-pnas-2011,zhang2012ordering},
where it is in general complicated to avoid interfering with the
phenomenon at study (see also~\cite{tong-jrsi-2022} for a more in-depth review).
Because of this, the role of fluctuations around the average
behaviors observed in crowd flows are rarely studied
\cite{moussaid-plos-2012,bongiorno-nat-2021}.

\begin{figure*}
  \centering
  \includegraphics[width=.99\textwidth]{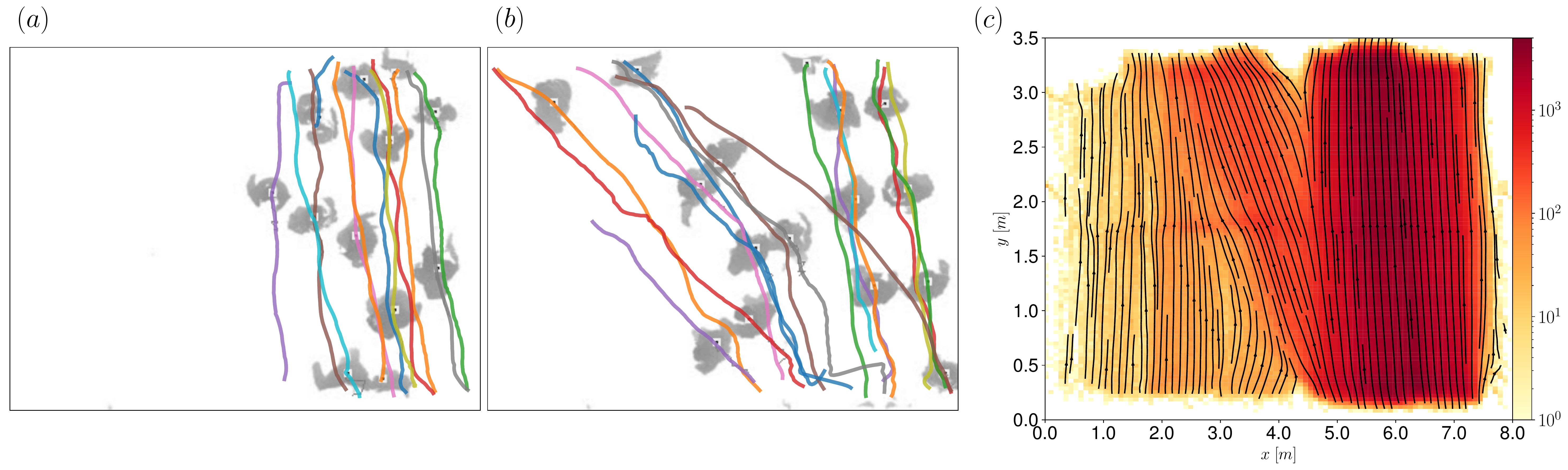}
  \caption{ Overview of the experimental data. In (a) and (b)
            we show two examples of overhead frames recorded by the depth cameras.
            The depth field in (a) depicts a set of $N = N_{\rm A} = 13$ 
            pedestrians all taking the straight path (path A), while in 
            (b) we provide an example of a more balanced pedestrian distribution.
            The gray shades represent the distance between each pixel 
            and the camera plane (i.e. the elevation from the ground).
            This type of data allows reliable pedestrian tracking 
            (see ``Materials and Methods'' for details).
            The automatic tracking output is overlayed as solid colored lines.
            (c) Heat-map of pedestrians position from the entire dataset. 
            We remark that the colorbar is given in logarithmic scale. 
            The streamlines of the (spatially binned) mean velocity vector
            are used in order to provide a visual representation of the most probable trajectories.
  }\label{fig:snapshot}
\end{figure*}

\begin{figure}
  \centering
  \includegraphics[width=.99\columnwidth]{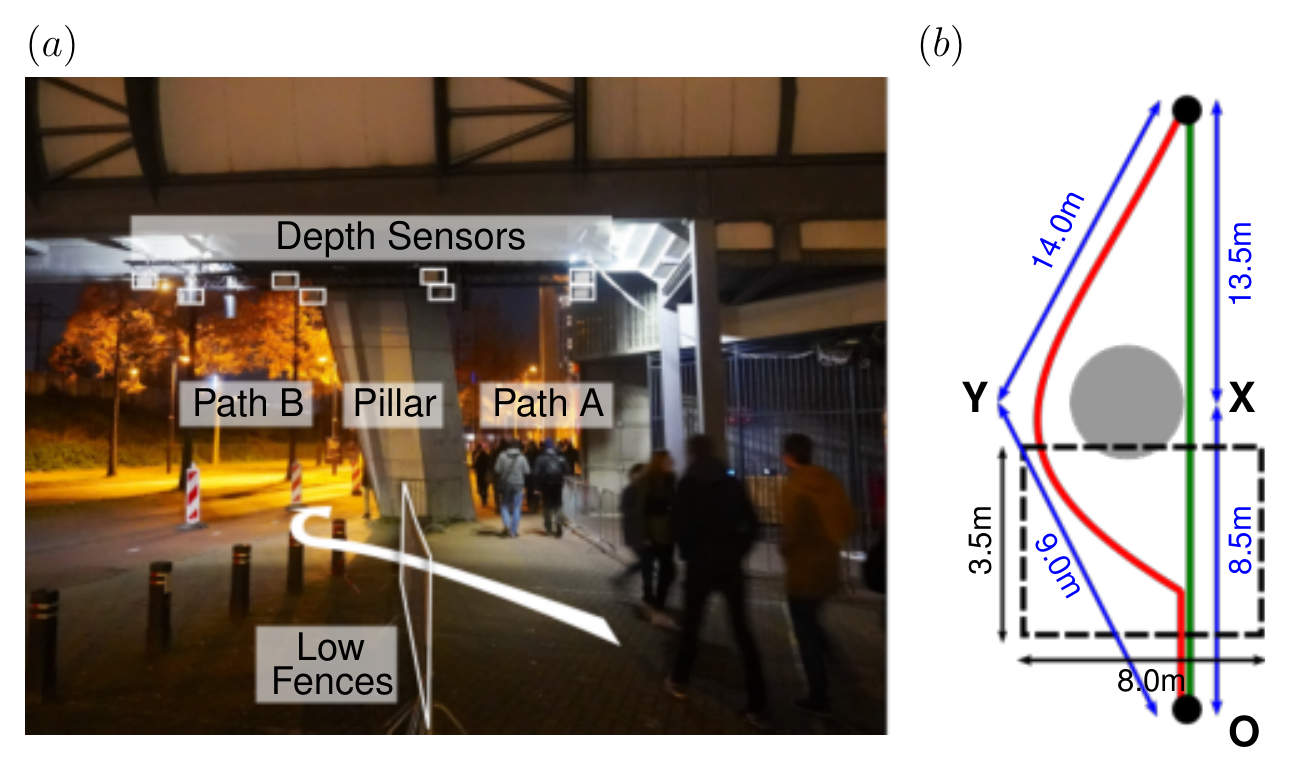}
  \caption{ Experimental setup from the viewpoint of a pedestrian
            walking towards the path bifurcation (a) and sketch of the floor
            plan (b).  A low fence blockage drives the pedestrian flow towards
            one same entrance point, while a set of bollards separates the
            bicycle lane from the adjacent road preventing pedestrians from
            entering the system from other locations or to exit by an area not
            covered by cameras.  A grid of 4x2 Orbbec depth cameras, hanging
            below the overpass connecting the Philips stadium to a nearby
            train station, is used to collect trajectories within the area
            marked by dotted black lines in (b).
  }\label{fig:experimental-setup}
\end{figure}

In this work, we analyze uni-directional pedestrian dynamics 
around a non-symmetric route bifurcation (Fig.~\ref{fig:experimental-setup}), 
as a paradigm scenario for non-trivial macroscopic routing. 
We base our analysis on high-accuracy high-statistics individual 
trajectory data collected during a week-long festival in Eindhoven (The Netherlands),
via overhead depth sensing (see Fig.~\ref{fig:snapshot} for an example), 
a methodology which has emerged in the last decade~\cite{brscic2013person,seer2014kinects,corbetta2014TRP,willems-sr-2020}
as an effective option to gather accurate tracking data in real-life, even at
high pedestrian density~\cite{kroneman-cd-2020}, while fully
respecting individual privacy. 
This approach enables arbitrarily long tracking campaigns during normal operations
of public facilities, and has allowed the analysis of fluctuations 
and rare events in pedestrian dynamics (e.g.~\cite{corbetta-pre-2017,corbetta-pre-2018,Brscic201477}).

We study the dynamics around the obstacle in
Fig.~\ref{fig:experimental-setup} for different density levels
by analyzing the trajectories of about $100.000$ individuals.
We focus on the statistics of collective routing
decisions in dependence on the local crowd density, $\rho$, here
considered via the instantaneous number, $N$, of pedestrians in the
facility.
In what follows we use these two quantities interchangeably, as they
can be put in relationship via $\rho = N / A_{\rm ref}$, where $A_{\rm ref} \approx 15.0~\rm{m^2}$
is the reference area effectively used by the pedestrians (see supplementary material).

Under these settings, we show that experimental observations are compatible
with realizations of a random process in which the crowd arranges in
such a way that the average (estimated) transversal time 
performs optimally with respect to all other traffic arrangements. 
In spite of the simplicity of the experimental
setup, the observed traffic departs from
a global optimal, transport-wise, partition, as the pedestrian throughput is
not maximized.

%=================================================================
%=================================================================

%=================================================================
%=================================================================

%=================================================================
\section{Measurement campaign}\label{sec:exp}
%=================================================================
%
We collected the trajectories used in the analysis presented in 
this paper during the GLOW light festival, in Eindhoven (The Netherlands), between
November 9th and 16th 2019.
The festival comprises a city-wide circular route, with mostly uni-directional
traffic. We established our measurement setup along the outer
perimeter of the Philips Stadium, few hundreds meters upstream and
downstream from the festival's light exhibitions. Pedestrians approaching the setup
faced the non-symmetric binary choice of bypassing, on either side, a
large support pillar (sustaining the stadium grandstands,
Fig.~\ref{fig:experimental-setup}(a)). On the right-hand side, the
path, from now on referred to as path A, was approximately straight, 
with free sight of the horizon. The longer path on the left-hand side, path B,
partially overlapping a bike lane (partially reserved to pedestrians), 
was rather curved around and following the pillar base 
(cf.~Figures~\ref{fig:experimental-setup}(a,b)).  
The crowd traffic in the area was stemmed by two types of barriers: 
several bollards placed on the side of path B separated the bicycle lane 
from the adjacent road, while a low fence directed the flow towards 
the path bifurcation from a single arrival basin. 

The geometrical definition of the length of the two paths,
respectively, $L_{\rm B}$ and $L_{\rm A}$,
is subject to a certain degree of arbitrariness, depending on
where the initial and final destination points are taken,
and on the considered connected trajectories. 
We shall characterize the geometry of our setup 
via the non-dimensional constant
\begin{equation}\label{eq:lambda}
  \lambda_g = \frac{L_{\rm B}}{L_{\rm A}} > 1,
\end{equation}
i.e. the ratio between the two paths lengths.

%=================================================================
\begin{figure}
  \centering
  \includegraphics[width=.99\columnwidth]{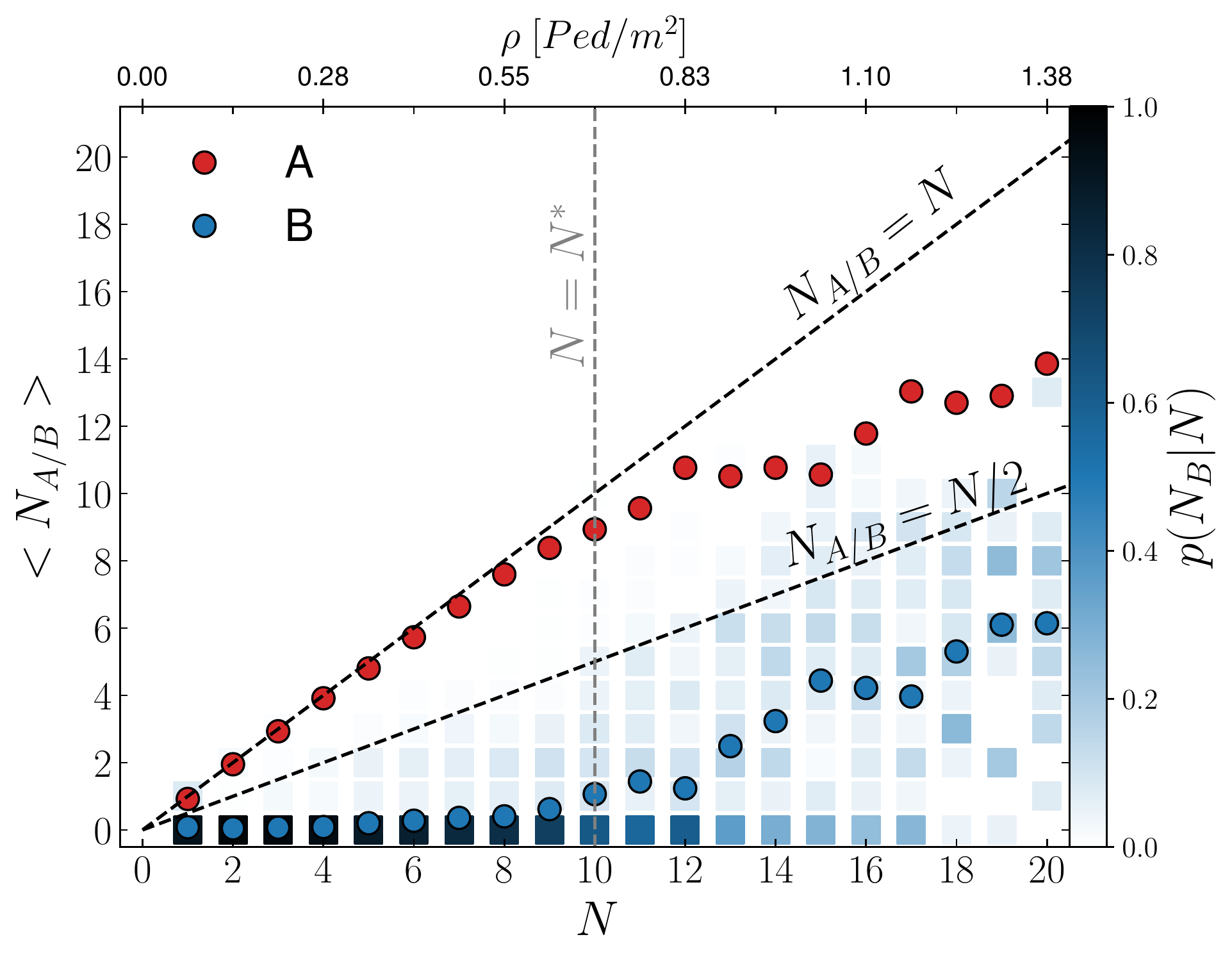}
  \caption{ 
            Average number of people taking path A ($\avg{N_{\rm A}}$, red dots) 
            and B ($\avg{N_{\rm B}}$, blue dots) as a function of the global pedestrian count $N$. 
            We observe that until $N < N^{*} = 10$, on average less than one person
            opts to travel along path B. Above the $N^{*}$ threshold people
            start making systematic use of path B and both diagrams exhibit
            a clear change.
            The blue colorbars provide a visual representation of the 
            probability distribution $P(N_{\rm B}|N)$ of number of people taking path B, 
            conditioned to the global pedestrian count $N$.
            Even when $N > N^{*}$ configurations in which no pedestrian walks
            on path B are frequent.}\label{fig:Nab-vs-N}
\end{figure}
%=================================================================
%
In order to provide an estimate for $\lambda_g$,
we consider two different approaches.
In the first one, we consider the right-triangle OXY in Fig.~\ref{fig:experimental-setup}(b),
with vertexes defined by the path midpoint at the entrance of the setup, 
right at the end of the low fences blockage (``O''), and the midpoints of paths A (``X'') and B (``Y'') across the pillar.
In this case it holds $\lambda_g \approx 1.06$.
If we restrict ourselves to the area covered by the depth sensors, 
we can also define $\lambda_g$ as a ratio between the length of 
a typical trajectory in B and in A (cf. Fig.~\ref{fig:snapshot}(c) 
which provides an overview of the trajectory data as a heat-map 
of pedestrian positions).
Including the uncertainty in the definition of these typical trajectories, 
it holds $1.3 \lesssim \lambda_g  \lesssim 1.4$.
We shall come back later to the analysis of $\lambda_g$ and on how it is 
perceived by single individuals.

In low density conditions, pedestrians opt for path A in the greatest majority of cases
(e.g. for $N < 10$ path A is preferred in $\approx 95 \%$ of cases).
This is shown in Fig.~\ref{fig:Nab-vs-N}, where we report the 
local average occupancy of the two paths, respectively $\avg{N_{\rm A}(N)}$ 
and $\avg{N_{\rm B}(N)}$, calculated on uncorrelated frames
as a function of the instantaneous count $N$ (see supplementary material).
As the number of pedestrians increases, we observe that path B ``activates'' 
as people start to systematically opt for it. 
We denote with $N^{*}$ the global pedestrians count at which path B activates, 
which we define as the minimum value of $N$ at which, on average, at least one
person takes path B; in our setup $N^{*} = 10$.

The local occupancy of paths A and B exhibits clear slope changes
around $N^{*}$.  In flow terms, $N^{*}$ corresponds to the transition
from a strongly unbalanced distribution, in which rarely a pedestrian
is found walking along path B, towards a more balanced A--B load
partition.

Figure~\ref{fig:Nab-vs-N} includes a visual representation of the
conditioned probability of the occupancy of path B, given the global
pedestrian count $N$, i.e.~$P(N_{\rm B} | N)$. Even when $N$ is much larger than
$N^{*}$, $P(N_{\rm B} | N)$ is bi-modal: path B remains often empty. 
For instance, at $N = 20$ we observe that in about $10 \%$ of the cases
pedestrians choose to walk only along path A.
This observation points to the presence of a collective dynamics in
which pedestrians at times follow others rather
than attempting to optimize the flow partitioning. 
So, how do pedestrian choose the path?
A quantitative modeling of this peculiar aspect will be the focus of our
analysis in the coming sections.

\begin{figure}
  \centering
  \includegraphics[width=.99\columnwidth]{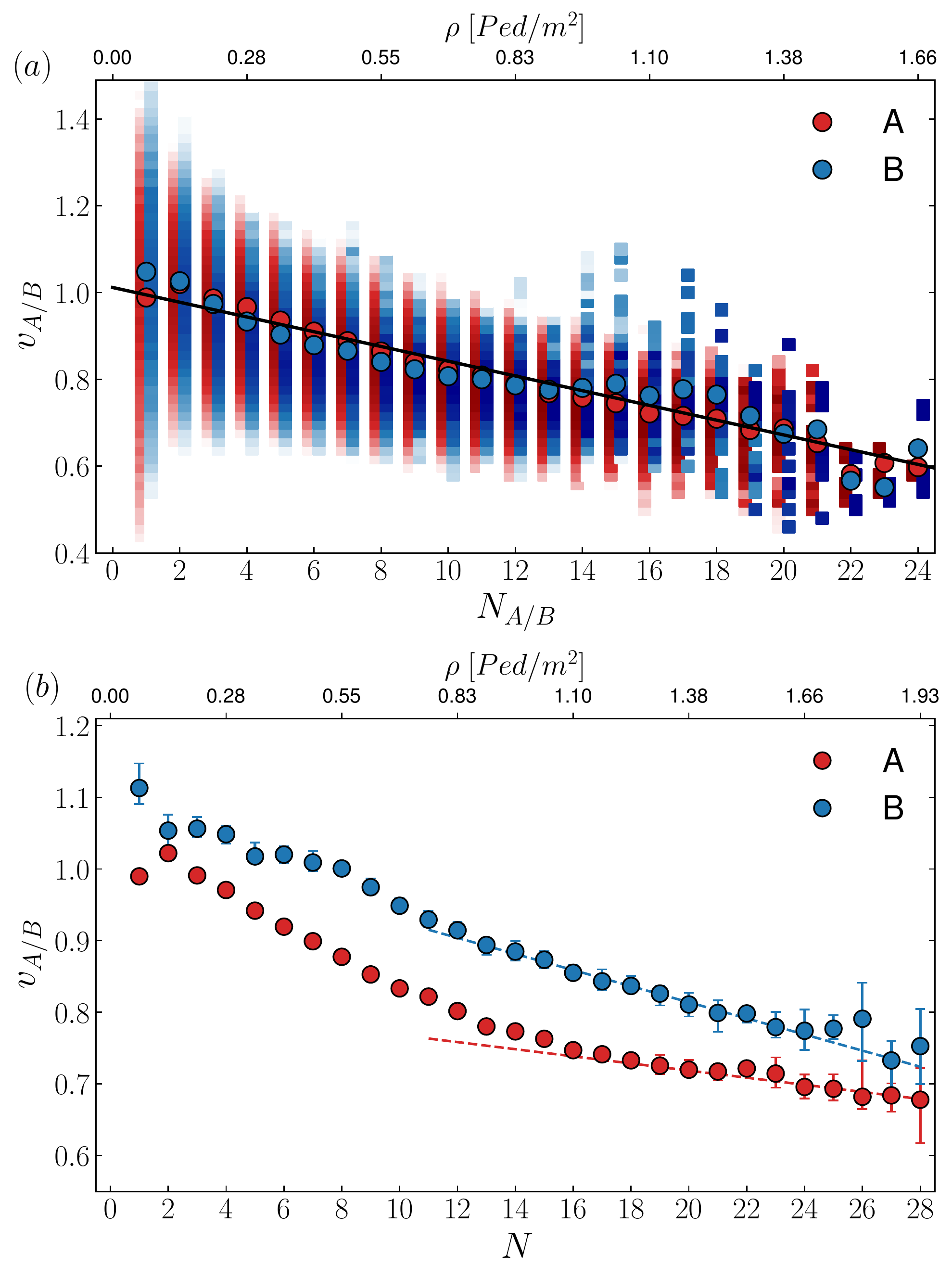} 
  \caption{ Fundamental velocity diagrams.
            (a) Local velocity as a function of the number of 
            people present along path A (red) and path B (blue).
            Dots represent the average values, while colorbars
            synthesize the probability distribution functions.
            The black solid line provides a linear fit of the local velocity
            diagrams, under the reasonable assumption that the same
            fundamental diagram applies to both paths.
            (b) Global velocity as function of the total number
            of people in the system. We highlight an evident change
            in the slope of the diagram for both path A and path B
            for $N > N^{*}$, which we model (dotted lines)
            with a re-parametrization of the local velocity diagram.
            (see main text for details).
          }\label{fig:velocity-fd}
\end{figure}

Different global and local pedestrian count levels (i.e.~in either path A or B)
reflect on different average walking
velocities. Fig.~\ref{fig:velocity-fd}(a) reports the (average) local
walking velocity along paths A and B as a function of the local
pedestrian count: $v_{J} = v_{J}(N_{J})$, with $J\in\{A,B\}$. In turn,
Fig.~\ref{fig:velocity-fd}(b) reports how velocity depends on the
global pedestrian count. These correspondences between
velocity and the density/pedestrians-count, generally
dubbed fundamental diagrams, are the most commonly adopted tool
for macroscopic descriptions of vehicular and pedestrian traffic
(cf.~e.g.~\cite{vanumu-etrr-2017,
  seyfried2005fundamental,jelic-pre-2012,bosina-strc-2018}).

%=================================================================
%=================================================================

As the number of pedestrians increases, the average walking velocity decreases.
Consistently with studies conducted in comparatively low-density
regimes~\cite{seyfried2005fundamental}, we observe, on average, a
linear decay trend in the local fundamental diagrams:
\begin{equation}\label{eq:local-fd}
  \avg{v_{J}(N_{J})} = v_0 - \kappa ~ N_{J},
\end{equation}
where $v_0$ is the ``free-stream velocity'' in the zero-density limit
and $\kappa$ fixes the diagram slope.
We assume the local fundamental diagram to be the same,
for people walking in path A and B. We have verified this by
performing a fit for the parameters $v_0$ and $\kappa$, independently,
for the two sets of pedestrians walking either of the two paths and
observing no significant differences.
In Fig.~\ref{fig:velocity-fd}(a), we show with a solid line the
best fit on the overall dataset, given by: 
$v_0 \approx 1.012~\rm{m/s}$, $\kappa \approx 0.017~\rm{m/s \cdot 1/ped}$, 
with the coefficient of determination $R^2 \approx 0.93$. 
Fig.~\ref{fig:velocity-fd}(a) additionally
reports the full conditioned probabilities $P(v_{J}|N_{J})$ that
highlights velocity fluctuations, $\epsilon$, around the average. 
We shall address these as independent with respect to the 
pedestrian count $N$, and additive with respect to the average
velocity, in particular
\begin{equation}\label{eq:additive-gaussian}
  \epsilon \sim \mathcal{N}(\mu=0, \sigma=0.15),
\end{equation}
where $\mathcal{N}$ is the Gaussian distribution, and the 
variance $\sigma$ has been estimated from the experimental data 
(see supplementary information).
The global fundamental diagrams, $\avg{v_{J}}=\avg{v_{J}(N)}$
in Fig.~\ref{fig:velocity-fd}(b), contrarily to their local 
counterparts, display qualitative and quantitative differences 
between the routes. For any value of $N$, the average walking 
velocity in path B is higher than in A:
\begin{equation}
  \avg{v_{\rm B}(N)} > \avg{v_{\rm A}(N)},\quad \forall N.
\end{equation}
Second, we observe a change in slope, $\partial_N\avg{v_{J}(N)}$, around
$N \approx N^{*}$ (we employ the symbol $\partial_N$ for the
partial derivative $\partial/(\partial N)$). For $N < N^{*}$, the
global diagram for path A coincides with its correspondent local
diagram:
\begin{equation}
  \avg{v(N_{\rm A})}\approx \avg{v_{\rm A}(N_{\rm A})},\quad  N < N^{*}.
\end{equation}
This is natural since, in this range, $N_{\rm A}(N)\approx N$ holds
(Fig.~\ref{fig:Nab-vs-N}). On path B, the velocity as a function of
$N$ decreases linearly, yet at a smaller rate than $-\kappa$
(i.e.~$\partial_N \avg{v_{\rm B}(N)} < -\kappa$). When $N$ is small,
path B is rarely employed (cf.~ probability distribution function of
the local density $N_{\rm B}$ in Fig.~\ref{fig:Nab-vs-N}). This allows
pedestrians to easily walk at their preferred walking speed
(i.e.~the free stream velocity $v_0$).

Conversely, when $N > N^{*}$, the activation of path B yields
$N_{\rm A}(N) < N$. This reflects in the slower decay of $\avg{v_{\rm A}(N)}$ as $N$
increases in comparison with the local counterpart:
\begin{equation}\label{eq:partialNavgVAleq}
  \partial_N \avg{v_{\rm A}(N)} < -\kappa,\quad  N > N^{*}.
\end{equation}
We can reconstruct the global fundamental diagram 
from the local diagram by considering $N_{J} = N_{J}(N)$. This
yields 
\begin{equation}
  \partial_N \avg{v_{J}(N)} \approx - \kappa \partial_N N_{J}\big|_N,\quad  N > N^{*},
\end{equation}
which satisfies \eqref{eq:partialNavgVAleq} since $\partial_N N_{J} < 1$
holds in the considered regime (cf.~Fig.~\ref{fig:Nab-vs-N}; see
the dotted lines included in Fig.~\ref{fig:velocity-fd}(b)).

%=================================================================
\begin{figure}
  \centering
  \includegraphics[width=.99\columnwidth]{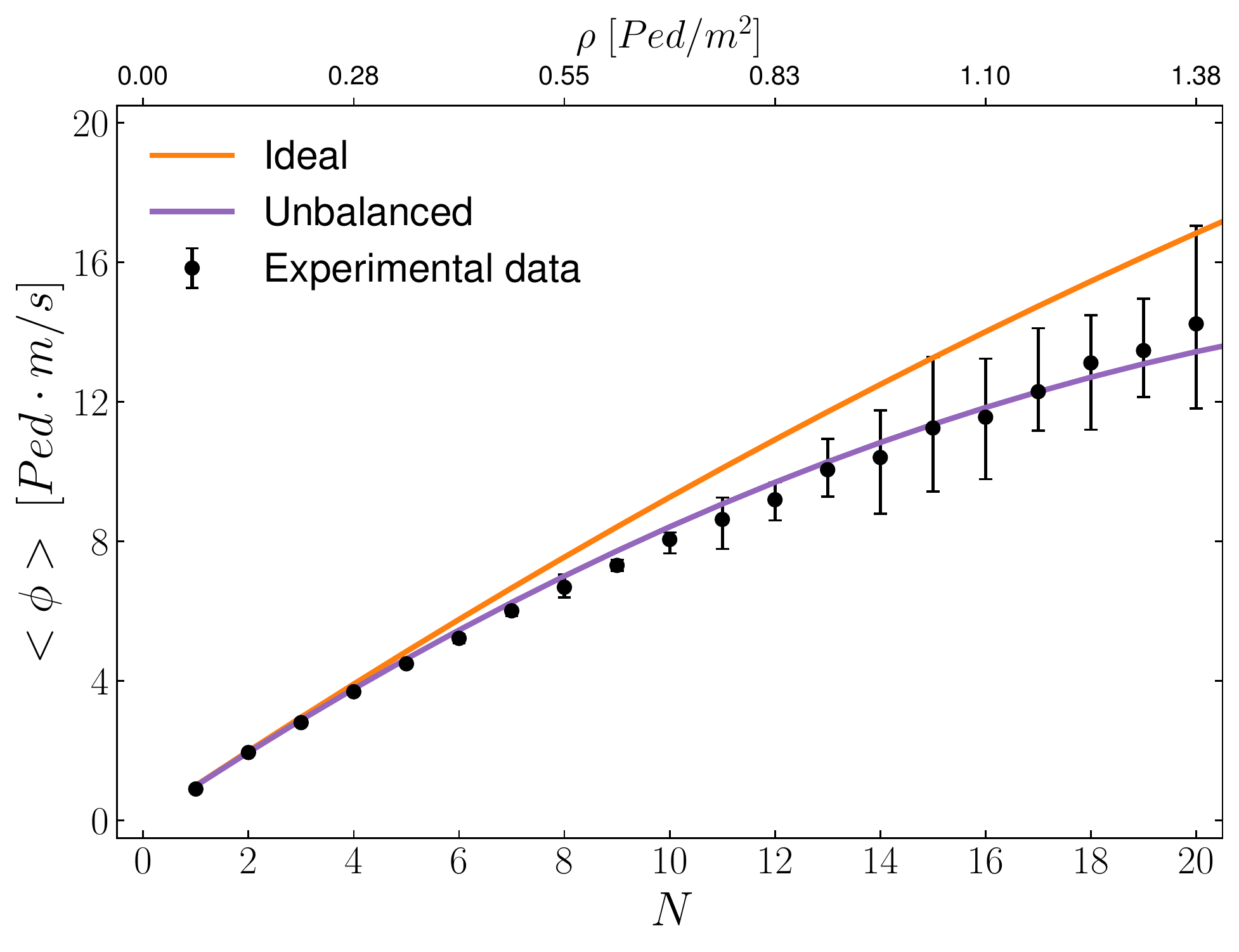}
  \caption{ 
            Average pedestrian flow (Eq.~\ref{eq:flow}) as a function of 
            the pedestrian count. The solid lines represent
            the theoretical maximum (orange color) and minimum (purple color) case scenarios.
            We observe that experimental results (black dots) on average
            closely follow the flow of the most unbalanced case.
            The error bars have been obtained by dividing the data into 10 bins,
            with the extrema of the error bars representing 
            the minimum and maximum average value per bin.
          }\label{fig:flow}
\end{figure}
%=================================================================

We conclude this section turning our analysis to the pedestrians flow,
which we define as:
\begin{equation}\label{eq:flow}
  \phi(N) = \avg{v_{\rm A}(N_{\rm A})} N_{\rm A}(N) + \avg{v_{\rm B}(N_{\rm B})} N_{\rm B}(N) .
\end{equation}
By making use of the fundamental velocity diagram, 
we can conveniently define a theoretical 
upper bound and lower bound for Eq.~\ref{eq:flow}, which are found 
respectively in correspondence of the optimal partitioning 
$N_{\rm A}(N) = N/2$, and the most unbalanced case $N_{\rm A}(N) = N$ (or likewise $N_{\rm A}(N) = 0$).
The above holds under the assumption that the section of path A 
equals that of path B, which is approximately true in our setup.
Combining this information with the velocity fundamental diagram
in Eq.~\ref{eq:local-fd}, we can define
\begin{equation}\label{eq:flow_model}
  \phi_{\rm ideal}(N) 
  = 
  v_{\rm A} \left(\frac{N}{2}\right) N , \quad \phi_{\rm unbalanced}(N) 
  = 
  v_{\rm A}(N) N .
\end{equation}
In Fig.~\ref{fig:flow} we compare the experimental data with the modeling
from Eq.~\ref{eq:flow_model}. 
The slope $\kappa$ determines the differences between the upper bound 
and lower bound, which in the density range considered are at most $20 \%$.
Nevertheless a clear trend emerges, with the experimental data closely 
following (on average) the flow of the highly unbalanced configuration;
this provides clear-cut evidence for pedestrians not managing to 
maximize the global throughput, despite the simplicity of the setup.

On these bases, in the following section we introduce
a model for studying the routing behavior and the features arising at the
transition around $N \approx N^{*}$, and where we assume that pedestrians
aim at optimizing their benefit (perceived travel time to destination).

%
%=================================================================
\begin{figure}
  \centering
  \includegraphics[width=.99\columnwidth]{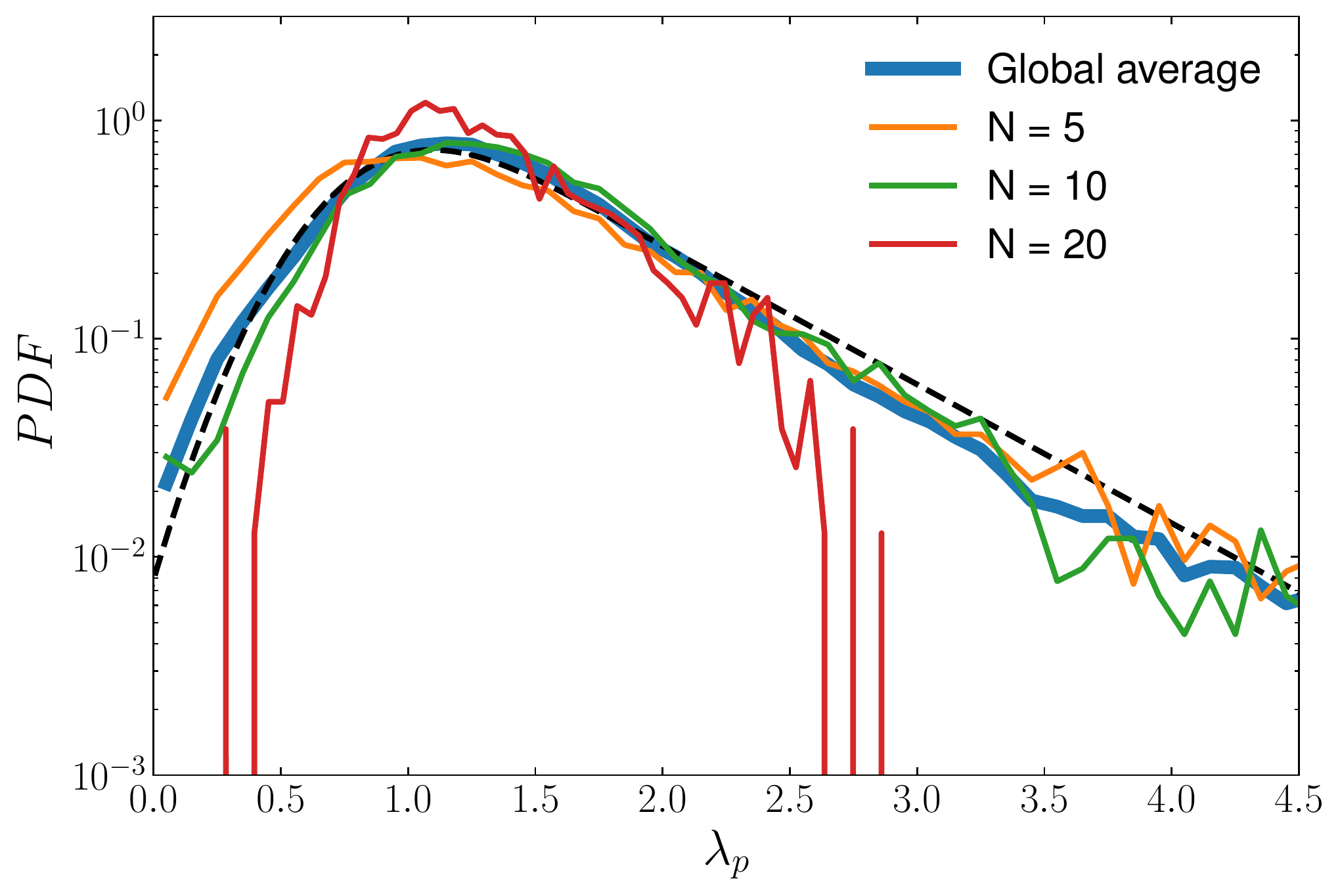}
  \caption{ Probability distribution function (PDF) of the perceived path
            length ratio  $\lambda_p$ (cf.~\eqref{eq:velocity_ratio}).            
            We report the distribution for for three different global 
            density values, respectively at low ($N = 5$), 
            intermediate ($N = 10$) and high ($ N = 20$) density values.
            The blue line shows the PDF obtained considering the
            overall dataset, with the black dotted line representing 
            a fit making use of an exponentially modified Gaussian
            (see Eq.~\ref{eq:lambdap-final}), with mean $\mu = 0.77$ 
            and standard deviation $\sigma= 0.30$, and an exponential 
            distribution with scale parameter $\beta = 0.68$.
          }\label{fig:vBovA}
\end{figure}
%=================================================================
%
%=================================================================
\section{Results}\label{sec:results}
%=================================================================
%
%=================================================================
\subsection{Model}\label{sec:model}
%=================================================================
%
We aim at a minimal model exposing the underlying mechanisms 
involved in the routing decision.

Although a time-dependent model for the probability
of choosing either paths, already pursued by the same authors~\cite{gabbana-cd-2021},
appears like a natural choice, its success is 
enslaved to the comprehension of the complex time correlation 
characterizing the choice process, or to phenomenological data-fitting~\cite{gabbana-cd-2021, wagoum-rsos-2017}.

The short duration of the festival, the relatively limited number of tracking hours, 
and the high variability in the crowd, make a time correlation analysis extremely 
challenging. 
Therefore, aiming at a bottom-up physical model, we pursue a 
time-independent approach.

We consider a simulated crowd of $N$ pedestrians
indexed by $i=1,\ldots,N$ about to cross the experiment area in
Fig.~\ref{fig:experimental-setup}. We allow each individual to
choose between path A or B in awareness of the choice of others. 
This gives configurations $c$ in the form of 
\begin{equation}
  c = (J^{(1)}, J^{(2)}, \ldots,J^{(N)}) ,
\end{equation}
where $J^{(i)}$ equals $A$ or $B$ depending on the path selected by
the the $i$-th pedestrian. 

Let $v^{(i)}_{J}=v^{(i)}_{J}(N_{J})$ be the walking velocity of the $i$-th
pedestrian on path $J$ as a function of the local density $N_{J}$,
i.e.~the local fundamental diagram (cf.~\eqref{eq:local-fd},
\eqref{eq:additive-gaussian}). We define the \textit{perceived} travel time
\begin{equation}
  \tau^{(i)}_{J} = g^{(i)}_{J}\left(\frac{L_{J}}{v^{(i)}_{J}(N_{J})}\right)
\end{equation}
in either paths to be a key variable in the A vs. B choice; 
here $g^{(i)}_{J}(\cdot)$ is a function mapping the actual ``geometric''
travel time $L_{J}/v^{(i)}_{J}$ to the perceived one.
The expression of $g^{(i)}_{J}$ will be discussed later on.

We consider a variational framework in which path choices are 
such that the minimum for the crowd-level functional
\begin{equation}\label{eq:overall-generic-loss}
  \loss = \loss(\tau^{(1)}_{J_1},\ldots,\tau^{(N)}_{J_N}), 
\end{equation}
is attained.
We consider a dynamics in which pedestrians arrange to reduce the
total \textit{perceived} travel time:
\begin{equation}\label{eq:avg-time}
  \loss = \sum_{i=1}^{N}\tau^{(i)}_{J_i}.
\end{equation}
Defining the discomfort functional $\loss$ is the modeling endeavor: the choice is not unique, 
yet \eqref{eq:avg-time} gave us the best agreement with observations; the interested reader
will find a comparison with a model adopting a different choice for $\loss$ 
in the supplementary information.

To summarize, we consider a system that takes the configuration
$c^*\in\Gamma$ for which 
\begin{align}\label{eq:complete-opt-probl}
  \min_{c\in\Gamma}\left[ \sum_{i=1}^{N}\tau^{(i)}_{J_i} \right] =
  \min_{c\in\Gamma}\left[ \sum_{i=1}^{N} g^{(i)}_{J}\left(\frac{L_{J}}{v^{(i)}_{J}(N_{J})}\right)\right], 
\end{align}
with $\Gamma$ representing the full set of $2^N$ distinct configurations,
and with the individual velocities (cf.~\eqref{eq:local-fd}) satisfying
\begin{equation}\label{eq:final-velocity-model}
  v^{(i)}(N_{J}) = v_0 - \kappa ~ N_{J} + \epsilon^{(i)},
\end{equation}
with $\epsilon^{(i)}$ independent and identically distributed realization
of \eqref{eq:additive-gaussian}.

Notably, the case $\epsilon^{(i)} = 0$, $g^{(i)}_{J}(x)=x$
(i.e.~deterministic velocity, and no fluctuations in the perception of the path-length) 
reduces to a Hughes-like model~\cite{hughes2003flow},
and has the analytic solution in terms of optical lenghts:
\begin{equation}\label{eq:deterministic-manifold}
  \frac{L_{\rm A}}{v_{\rm A}(N_{\rm A})} = \frac{1}{\sqrt{\lambda_g}} \frac{L_{\rm B}}{v_{\rm B}(N_{\rm B})} \quad ,
\end{equation} 
where we have dropped the index $i$ since pedestrians are now
indistinguishable from each others. 
The above implies the following expression for $N_{\rm A}=N_{\rm A}(N)$:
\begin{equation}\label{eq:variational-principle-solution}
  N_{\rm A}(N) = \min \left\{ N , \frac{\kappa N + v_0(\sqrt{\lambda_g} -1)}{\kappa ( \sqrt{\lambda_g} + 1 )}  \right\} \quad .
\end{equation} 
Moreover, from Eq.~\ref{eq:deterministic-manifold} we can define
a link between $\lambda_g$ and the local velocity of pedestrians in path A and B: 
$\lambda_g = \left( \frac{v_{\rm B}(N_{\rm B})}{v_{\rm A}(N_{\rm A})} \right)^2 $.
The above expression suggests an alternative pathway for measuring $\lambda_g$
directly from experimental data. To this aim, we introduce the
instantaneous quantity
\begin{equation}\label{eq:velocity_ratio}
  \lambda_p(t) = \left( \frac{\hat{v}_{\rm B}(t)}{\hat{v}_{\rm A}(t)} \right)^2,
\end{equation}
where $\hat{v}_{\rm B}$ (resp. $\hat{v}_{\rm A}(t)$) indicates the
average walking speed of pedestrians in path B (resp. A) measured at
time $t$.

In Fig.~\ref{fig:vBovA}, we show the probability distribution function (PDF)
of $\lambda_p$, for the overall dataset, and also conditioned on a few
selected values of $N$; we report three
representative examples at low, intermediate and large density values
(PDFs are restricted to meaningful cases $N_{\rm A},N_{\rm B}>0$).
Two aspects emerge.
The modal value, $\mbox{mode}(\lambda_p)\approx 1.2$, of the
distributions is independent on the global pedestrian count $N$,
consistently with the deterministic model in
\eqref{eq:deterministic-manifold}.
While, $\mbox{mode}(\lambda_p)$ is comparable with the estimates 
of $\lambda_g$ provided in the previous section, we observe that 
the distributions for $\lambda_p$ are skewed and carry heavy tails,
in particular at low densities.

These are due to observed configurations strongly departing from
the deterministic optimum in \eqref{eq:deterministic-manifold}.
Right tails corresponds to cases in which many pedestrians walk along
path A even though it might have been less costly (in $\loss$ terms) to take B.  
This can be motivated considering
that opting for path B involves traveling around an obstacle which hides the
horizon and to invade the (temporarily closed) bike lane.

The variance of the distributions decreases with the
global density.  This is consistent with the fact that for $N > N^{*}$
the load between A and B gets (on average) increasingly balanced,
conversely, the herding becomes weaker (see Fig.~\ref{fig:Nab-vs-N}).

In the next section, we compare Monte Carlo simulations 
of the dynamics considering various models for $\lambda_p$, which we integrate in
\eqref{eq:avg-time}-\eqref{eq:complete-opt-probl} by defining the
conversion functions $g^{(i)}_{J}(\cdot)$ as $i$-independent
(i.e.~pedestrian-independent) rescaling factors
\begin{align}
  &g_{A}\left(\frac{L_{\rm A}}{v_{A,i}(N_{\rm A})}\right) = \frac{L_{\rm A}}{v_{A,i}(N_{\rm A})} \nonumber \\
  &g_{B}\left(\frac{L_{\rm B}}{v_{B,i}(N_{\rm B})}\right) = \frac{\lambda_p}{\lambda_g}\frac{L_{\rm B}}{v_{B,i}(N_{\rm B})} 
                                                    = \lambda_p\frac{L_{\rm A}}{v_{B,i}(N_{\rm B})}.\label{eq:g-model}
\end{align}
Following the PDF in Fig.~\ref{fig:vBovA}, we fit $\lambda_p$ with
an $N$-independent exponentially modified Gaussian distribution
(i.e.~the sum of independent normal and exponential random variables):
\begin{equation}\label{eq:lambdap-final}
  \lambda_p = X + Y,
\end{equation}
where $X \sim \mathcal{N}(\mu = 0.77, \sigma= 0.30)$ and $Y \sim \mbox{Exp}(\beta = 0.68)$,
where $\beta$ is the scale parameter of the exponential distribution;
observe that the expected value is given by $E[X+Y] = \mu + \beta = 1.45$.

%=================================================================
\subsection{Numerical Results}\label{sec:numerics}
%=================================================================

While the deterministic version of the model offers access to a simple
analytic solution \eqref{eq:deterministic-manifold}, this is not the
case for the non-deterministic model
(Eqs.~(\ref{eq:complete-opt-probl}-\ref{eq:final-velocity-model}-\ref{eq:g-model}-\ref{eq:lambdap-final})).
Therefore, to perform our analysis and compare with measurements we
rely on Monte Carlo simulations to identify the statistics
of optimal configurations in dependence on the stochastic terms
considered:  $c^* = c^*(\lambda_p, \epsilon^{(1)},\ldots,\epsilon^{(N)})$.

\begin{figure}
  \centering
  \includegraphics[width=.99\columnwidth]{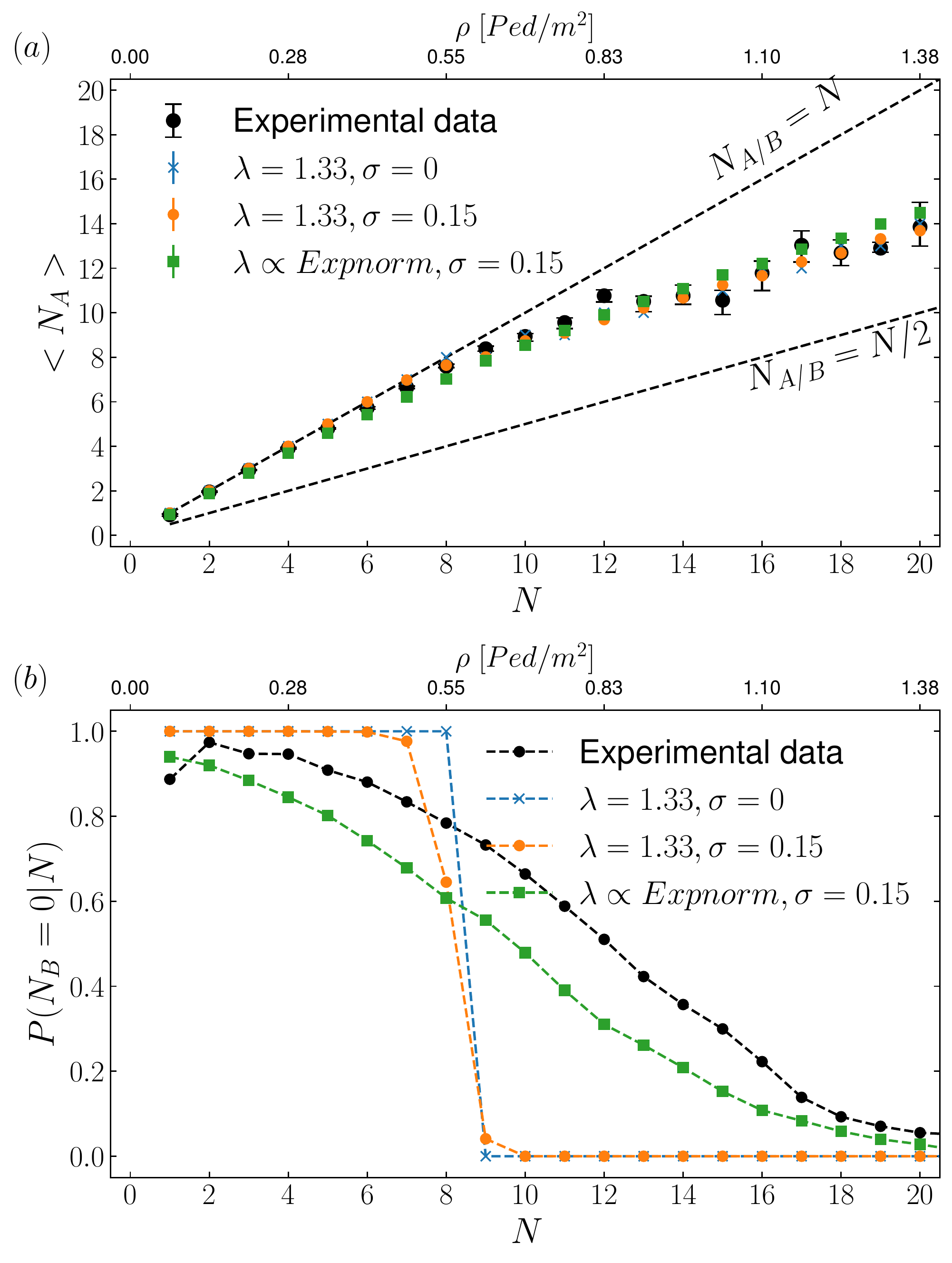}  
  \caption{ Comparison of numerical results from simulations 
            against experimental data.
            (a) $<N_{\rm A}(N)>$: the average number of people 
                taking path A as a function of the global 
                pedestrian count $N$.
            (b) $P(N_{\rm B} = 0 | N)$, the Bernoulli probability 
                of observing configurations in which 
                no pedestrians walk across path B,
                conditioned on the global pedestrian count $N$.
            It is evident that fluctuations on the perceived 
            path length allow a more realistic description 
            of the transition around $N^{*}$, as shown in b), 
            still correctly capturing the average behavior, 
            as shown in a).
          }\label{fig:num_res}
\end{figure}

In Fig.~\ref{fig:num_res}(a) we compare the model and experimental
data on the average number of people taking path A, $<N_{\rm A}>$, 
conditioned to the global density $N$.
The numerical results provide a good description of the
measurements, and, in particular, they capture the transition at
$N^{*}$.
%
%
%=================================================================
%=================================================================
%
The model is capable of reproducing, with very good accuracy, also the
footprints of the herding effect: this is shown
Fig.~\ref{fig:num_res}(b), reporting the (Bernoulli) probability of
observing exactly zero pedestrians walking along path B, conditioned
to $N$ (i.e. $P(N_{\rm B}=0|N)$).
In order to obtain a good agreement between experimental data and simulations
we have tuned the parameters of the distribution from which $\lambda_p$
is drawn; the results presented in this section make use of Eq.~\ref{eq:lambdap-final}
with $X \sim \mathcal{N}(\mu = 1.15, \sigma= 0.20)$ and $Y \sim \mbox{Exp}(\beta = 0.33)$.

With the aim of exposing the role of random fluctuations, in
Fig.~\ref{fig:num_res} we show the results obtained by
employing a fully deterministic model (i.e.~with a deterministic
fundamental velocity diagram, $\epsilon^{(i)}=0$, and with a constant
value for $\lambda_p$) as well as a case in which we allow
fluctuations in the velocity, but no stochasticity on $\lambda_p$.

The deterministic model well captures the average 
routing choice performed by pedestrians, as shown in Fig.~\ref{fig:num_res}(a).
On the other hand, it also highlights a sharp 
transition at $N^{*}$ (see Fig.~\ref{fig:num_res}(b)): when $N < N^{*}$
all pedestrians systematically route for path $\rm A$, while for 
$N > N^{*}$ the optimal configurations do not allow for cases 
in which exactly zero pedestrians are found walking along path B.

When including fluctuations in the velocity (orange curves) we obtain two
relevant effects connected to each other.
The walking speed variability creates (rare) optimal configurations $c^*$ 
with pedestrians on path B, even at density values $N < N^{*}$; this
effect, only slightly visible in Fig.~\ref{fig:num_res}(b), 
becomes more pronounced as the variance associated to $v_0$ is increased,
in turn leading to a smaller predicted value for $N^{*}$ .

\begin{figure*}
  \centering
  \includegraphics[width=.99\textwidth]{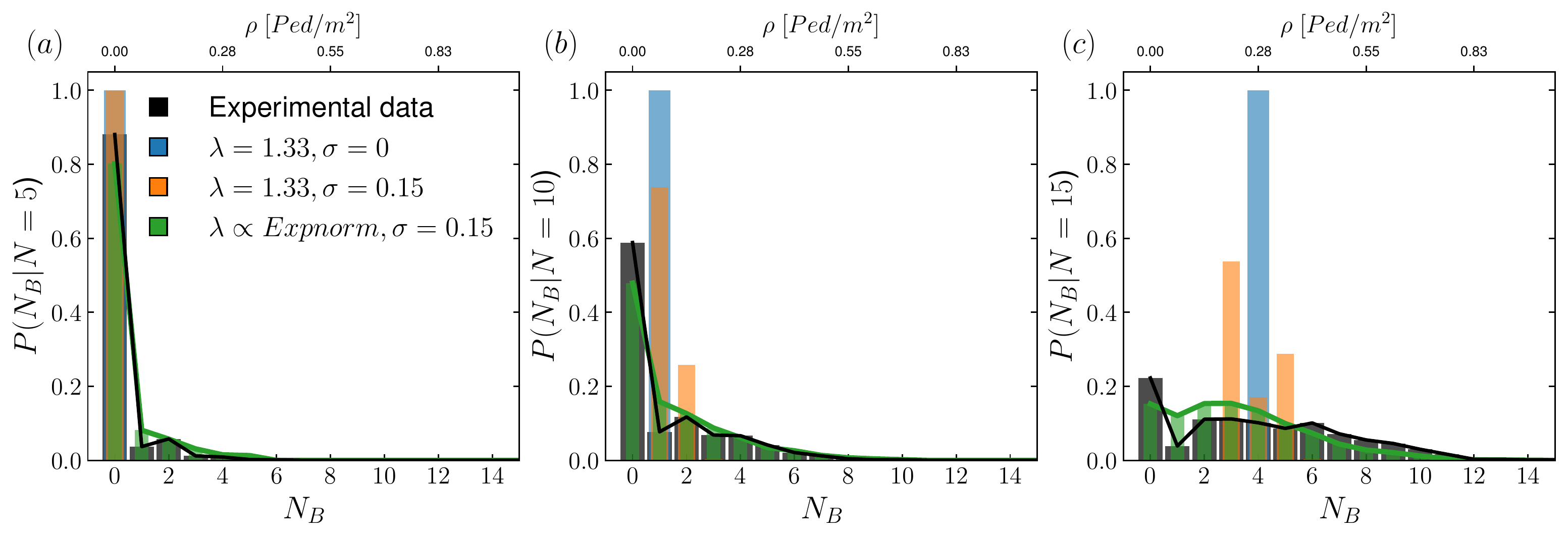}
  \caption{ Probability distribution function of number of people
            in path B, $N_{\rm B}$, conditioned on the global count $N$.  
            We report three examples, representative of three different
            levels of density: from left to right $N = 5, 10, 15$.
            The experimental data (black bars) are compared against
            the results obtained employing three different models:
            In blue the results obtained employing a deterministic model
            in which individual fluctuations are neglected,
            in orange the results of a stochastic model
            accounting for fluctuations in individuals free-stream velocities,
            and finally in green the results of a stochastic model
            accounting for fluctuations in both free-stream velocities
            and path length perception for single individuals.
          }\label{fig:pNB_cond_N}
\end{figure*}

Introducing fluctuations in the model
is crucial to provide an accurate description of 
the variability observed in the experimental data.
This is clearly shown in Fig.~\ref{fig:pNB_cond_N}, where we plot the
probability distribution function for the number of people walking
along path B, conditioned to the global count $N$.  The figure reports
three representative examples corresponding to different values of
$N$.
For low density values, the PDFs show a strong peak at $N_{\rm B} = 0$.  As
$N$ increases, the bins corresponding to $N_{\rm B} > 0 $ start populating
and, eventually, a bi-modal distribution emerges, together with an
increased variability in the observed configurations.

Comparing once again the numerical results with the experimental data
we can observe that the deterministic model cannot be used to describe
the variability observed in the data, although it can provide an
approximation to the average of PDFs.
While introducing fluctuations in the pedestrians velocity 
only slightly increases the variability of the PDFs for $N > N^{*}$,
it is only with the superposition of the herding effect (green curves)
that the model is able to provide a good description of the PDFs.
Remarkably, we are able to reproduce to good accuracy also the
spikes in correspondence of $N_{\rm B} = 0$ at large values of $N$.

In conclusion, we have shown that fluctuations are crucial for giving 
a realistic representation of the behaviors observed around $N \approx N^{*}$.

%=================================================================
\section{Discussion}\label{sec:discussion}
%=================================================================
%
%=================================================================
%=================================================================

In this work, we have exposed the crucial role played by individual variability
in pedestrians routing choices. Fluctuations emerge as a key element
in explaining (intermittent) transitions from highly unbalanced to
more balanced configurations which, on average, lead to a sub-optimal 
traffic partitioning.

We have based our analysis on a large dataset of pedestrian trajectories 
collected during an unprecedented high-accuracy pedestrian tracking campaign.
We have considered a simplified setup in which a unidirectional pedestrian flow 
is confronted with a binary choice between two paths, 
presenting marginal differences in terms of length and geometrical complexity. 
We regard this setup as an excellent prototype for more complex scenarios where,
e.g., the trajectory of a pedestrian results from the concatenation of 
multiple binary choices.

We have developed a time-independent variational model, which has
allowed to successfully describe, both at a qualitative and
quantitative level, the observed macroscopic patterns.  
Our modeling shows that we can explain the crowd behavior 
by considering a crowd-level minimization of the \textit{estimated traveling time}, 
and accounting for the inherent stochasticity of (i) the walking speed 
of each single pedestrian, and (ii) the estimation of the path length.

In spite of the simplicity of the experimental setup, our analysis
highlights a systematic deviation from global optimum configurations,
leading to the global pedestrian throughput not being maximized.
Additionally, further and sudden capacity drops appear due to the
occurrence of herding behaviors - in which the crowd blindly opt for
a highly sub-optimal ``follow the lead'' choice, rather than completely 
leveraging the allowed walking space.
We remark that in our analysis we use the word herding in a broad sense,
including both following effects as well as the presence of social groups
attending the event. This choice is due to the fact that groups cannot 
be easily identified in the relatively short-scales of the experiment 
presented in this work, something on the other hand possible when observing 
trajectories in a much larger space/time frame~\cite{pouw-plos-2020}.

These results clearly point towards the necessity of implementing
efficient crowd management measures in order to increase comfort and safety,
based on a deeper understanding of the physics of crowds.

To conclude, in this work we have introduced an approach 
for analyzing the statistics and the efficiency of macroscopic crowd configurations, 
highlighting an intrinsic sub-optimality in the natural flow of pedestrians, 
while setting a standard for effective quantitative modeling.

%=================================================================
\section*{Acknowledgments}
%=================================================================
We acknowledge Philips Stadion, TU/e Intelligent Lighting Institute 
and Signify for their support,
and Cas Pouw for his help in the data acquisition process.
This work is partially supported by the HTSM research 
programme "HTCrowd: a high-tech platform for human 
crowd flows monitoring, modeling and nudging" with project 
number 17962 and partially by  the VENI-AES research programme 
"Understanding and controlling the flow of human crowds" 
with project number 16771, 
both financed by the Dutch Research Council (NWO).
%=================================================================

%=================================================================
\section*{Data availability}
%=================================================================
The dataset with the pedestrian trajectories used in our 
analysis is available at \url{https://doi.org/10.5281/zenodo.7007358}, 
whereas examples and processing scripts can be found at 
\url{https://github.com/crowdflowTUe/2022_fluctuations_in_routing_glow}.

\bibliography{master.bib}

%%%%%%%%%% Merge with supplemental materials %%%%%%%%%%
\pagebreak
\widetext
\begin{center}
\textbf{\large Supplementary Information for \\``Fluctuations in pedestrian dynamics routing choices''}
\end{center}
%%%%%%%%%% Merge with supplemental materials %%%%%%%%%%
%%%%%%%%%% Prefix a "S" to all equations, figures, tables and reset the counter %%%%%%%%%%
\setcounter{equation}{0}
\setcounter{figure}{0}
\setcounter{table}{0}
\setcounter{page}{1}
\makeatletter
\renewcommand{\theequation}{S\arabic{equation}}
\renewcommand{\thefigure}{S\arabic{figure}}
\renewcommand{\thetable}{S\arabic{table}}
% \renewcommand{\bibnumfmt}[1]{[S#1]}
% \renewcommand{\citenumfont}[1]{S#1}
%%%%%%%%%% Prefix a "S" to all equations, figures, tables and reset the counter %%%%%%%%%%

%=================================================================
\section*{Experimental Setup}
%=================================================================

The trajectories used in the analysis presented in this work
have been collected during the 2019 edition of the GLOW light 
festival in Eindhoven (The Netherlands).
The experiment lasted the entire duration of Glow 2019, 
from November 9th until November 16th, 2019. 
The tracking was performed during the festival opening hours,
every day from 18:00 until 00:00.
The data collected on the 14th of November has not been included
in the analysis, since on that day the experimental setup was
modified in order to evaluate the impact of changing the lighting 
conditions on the crowd dynamic.

\begin{figure}
  \centering
  \includegraphics[width=.49\textwidth]{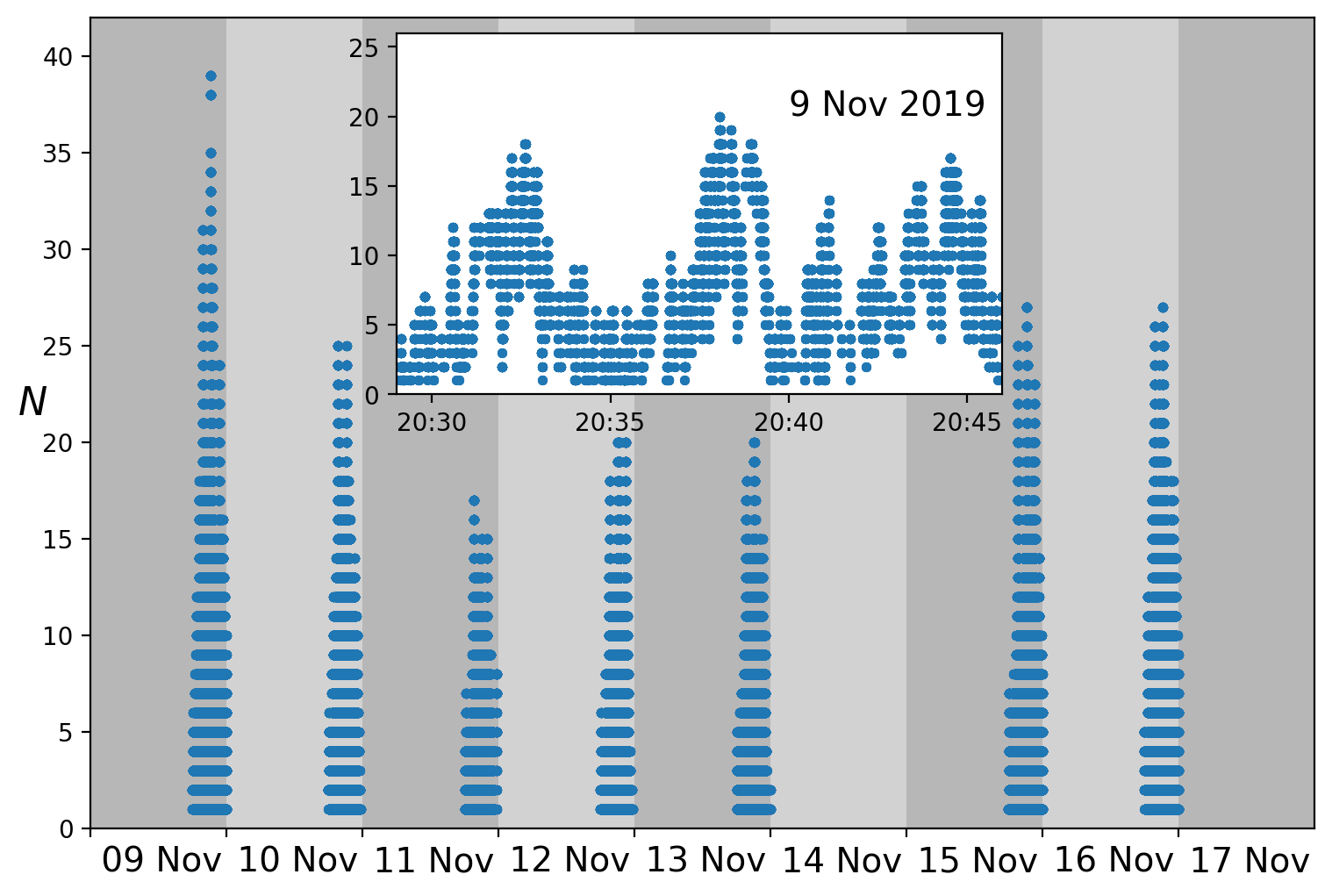}
  \caption{ 
            Pedestrian global count, $N$,  as a function of time
            during the 8 days of the festival. The inset shows the fluctuations
            in the number of pedestrians observed in a 15 minute window around 20.40 
            during the first day of the experiment, highlighting the presence
            of both low and high density scenarios. The data collected on the 14th of November
            has not been used in our analysis since the experimental setup had been
            employed in a modified form and with a different purpose.
          }\label{fig:density-vs-time}
\end{figure}

In Fig.~\ref{fig:density-vs-time} we show the pedestrian count
as a function of time, with a inset highlighting the fluctuations
in the number of pedestrians observed in a 15 minute window.

%=================================================================
%=================================================================

%=================================================================
\section*{Pedestrian Sensing}
%=================================================================

We collected raw depth images of a walkable area of about $30\,m^2$ 
via 8 Orbbec Persee sensors attached underneath a pedestrian overpass, 
and arranged in a 4x2 grid (see Fig.~2 in the main text).
The depth cameras acquired images at a frame rate of $30\,$Hz.  
The trajectories have been obtained from the raw depth images 
via the Height-Augmented Histogram of Oriented Gradients algorithm (HA-HOG)
(see~\cite{kroneman-cd-2020} and ~\cite{corbetta-pre-2018,corbetta-CDA34} 
for detailed explanations on pedestrian tracking via depth sensor grids).

\begin{figure}
  \centering
  \includegraphics[width=.49\textwidth]{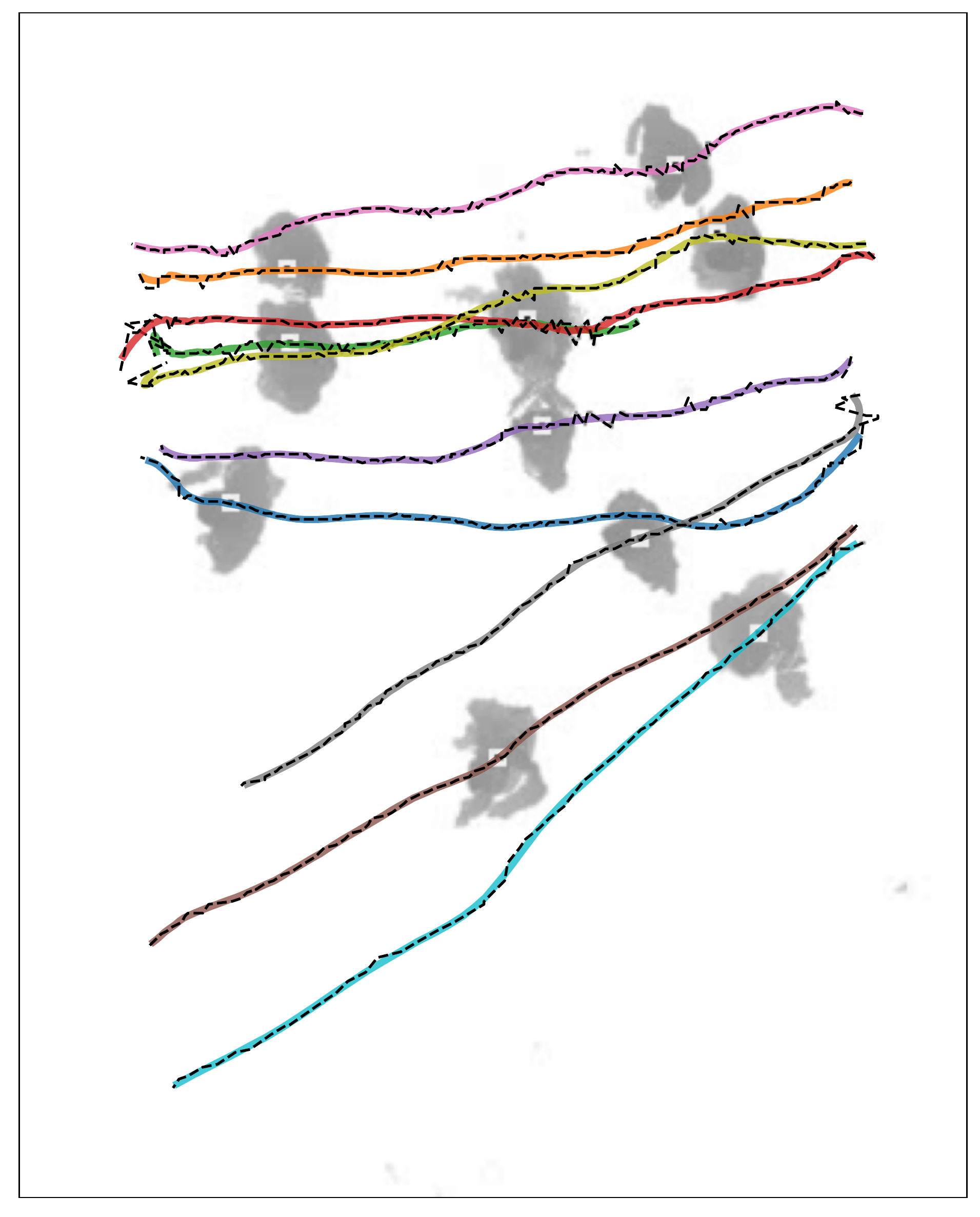}
  \caption{ Snapshot of data recorder by the depth cameras.
            The black dotted lines represent the trajectories of each pedestrian,
            obtained by applying the HA-HOG algorithm.
            Solid lines have been obtained by applying a 
            Savitzky-Golay filter.
          }\label{fig:snap}
\end{figure}

In Fig.~\ref{fig:snap} we show a depth map example with the trajectories 
resulting from the tracking of the 10 pedestrians in overlay.

The black dotted lines represent the raw trajectories,
obtained by applying the HA-HOG algorithm.
Solid lines represent the result of applying a Savitzky-Golay filter~\cite{savitzky-ac-1964}
to the trajectories. This operation allows to reduce the level of
noise and discontinuities which affect the calculation of
derivatives, used, for example, to estimate the instantaneous velocity
of pedestrians:
\begin{equation}
  \bm{v}(t) = \left( \frac{ \tilde{s}_x(t + \Delta t) - \tilde{s}_x(t)}{\Delta t}, 
                     \frac{ \tilde{s}_y(t + \Delta t) - \tilde{s}_y(t)}{\Delta t} 
              \right)
\end{equation}
where $\tilde{ \bm{s} }(t)$ represents the spatial position within the
trajectory of a pedestrian at time $t$ after having applied the 
Savitzky-Golay filter, and $\Delta t = 1 / 30 \rm{s}$ follows from the
cameras aquisition rate.

%=================================================================
\section*{Data analysis}
%=================================================================

\paragraph*{Refinement of the dataset}
In the analysis we have considered only configurations 
from uni-directional flows. We have dropped all trajectories of pedestrians
traveling in the opposite direction with 
respect to the viewpoint in Fig.~2 in the main text,
as well as all trajectories interacting directly or indirectly with them,
i.e. both being present at the same time in at least one frame,
or sharing a frame with a pedestrian who has previously interacted with
a trajectory going in the opposite direction.

Although the bike lane adjacent to the experimental setup was 
supposedly closed to traffic during the festival hours, 
cyclists and runners were still present and able to access it
from the street.
In order to drop from the dataset cyclists, runners, 
as well as people standing still under the area covered by our sensors,
we have retained trajectories with instantaneous velocities $v(t)$
in the interval of $[0.05, 2.9]~\rm{m/s}$
and average velocity $\langle v \rangle$ of $[0.15, 1.5]~\rm{m/s}$.

Throughout the week, we have collected $192,229$
individual trajectories. Following the above discussion, 
the analysis retains $101,867$ among these trajectories.

%=================================================================
%=================================================================

\paragraph*{Data de-correlation}
The calculation of the average number of pedestrians walking 
in each path as a function of the global pedestrian count requires
extra care due to presence of strong correlations between 
consecutive frames.

For this reason, in our analysis we have taken into consideration
only configurations at least $4$ seconds apart from each other.
This value is larger than the average time duration of 
a single trajectory, which in our data corresponds to $\approx 3.9$ seconds.

\begin{figure}
  \centering
  \includegraphics[width=.49\textwidth]{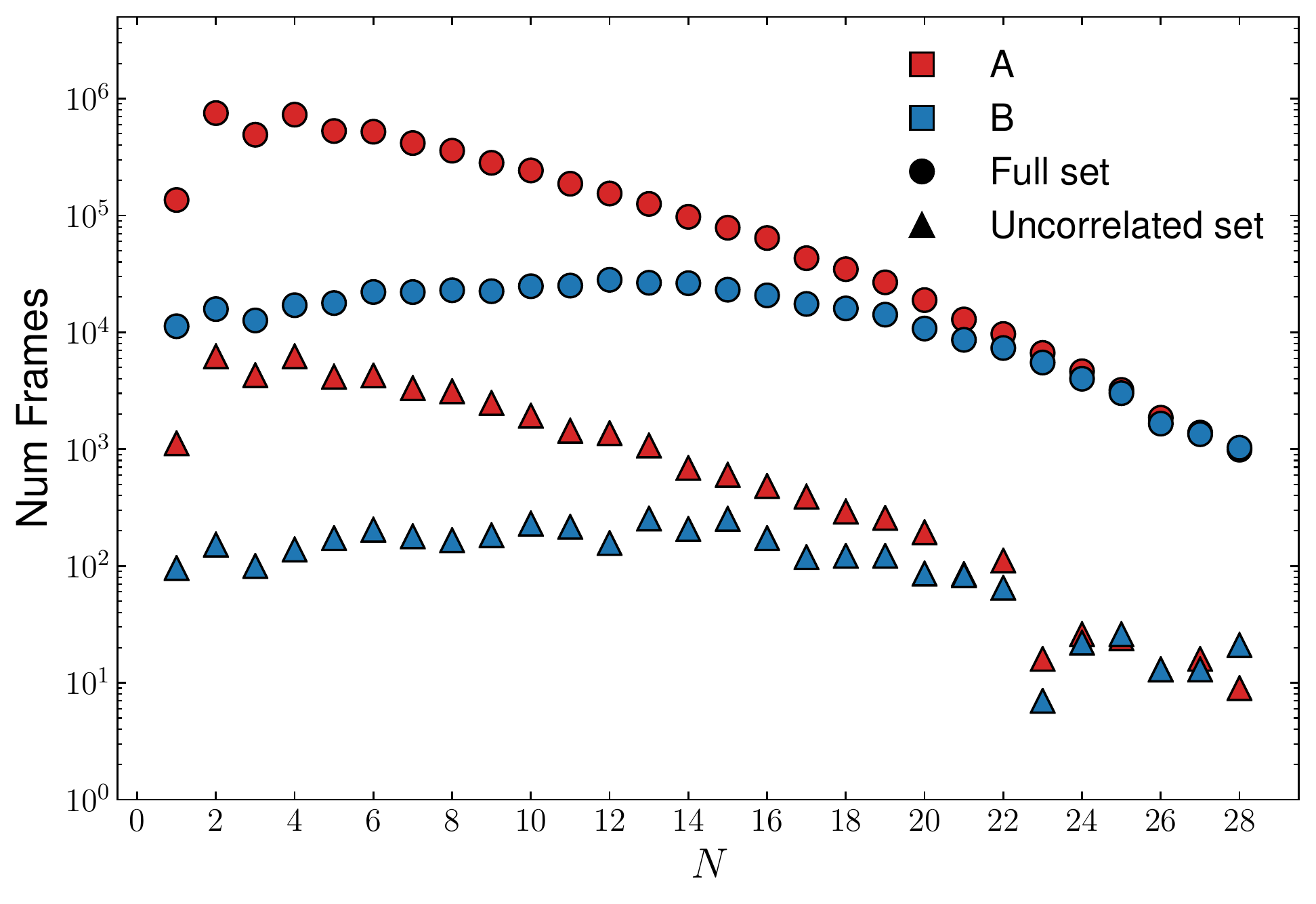}
  \caption{ 
            Number of frames contributing
            to the statistics of pedestrians walking in path A and B as a function
            of the global pedestrian count. Dots represent the full dataset,
            which we use to compute statistics on (and related to) the velocity of pedestrians.
            Triangles represent the dataset consisting of uncorrelated configurations, 
            taken at least $4$ seconds apart from each other, which we use
            to compute statistics on (and related to) the pedestrian count.
            The plot gives an indication of the statistical resolution of
            our analysis.
          }\label{fig:nframes}
\end{figure}

In Fig.~\ref{fig:nframes} we show the number of frames contributing
to the statistics of pedestrians walking in path A and B as a function
of the global pedestrian count. Dots represent the full dataset,
whereas triangles represent the uncorrelated dataset.

For the calculation of the local and global velocity fundamental diagram
we have made use of the full dataset, since, in this case, 
time correlations do not introduce biases in the analysis.

%=================================================================
%=================================================================

\begin{figure}
  \centering
  \includegraphics[width=.49\textwidth]{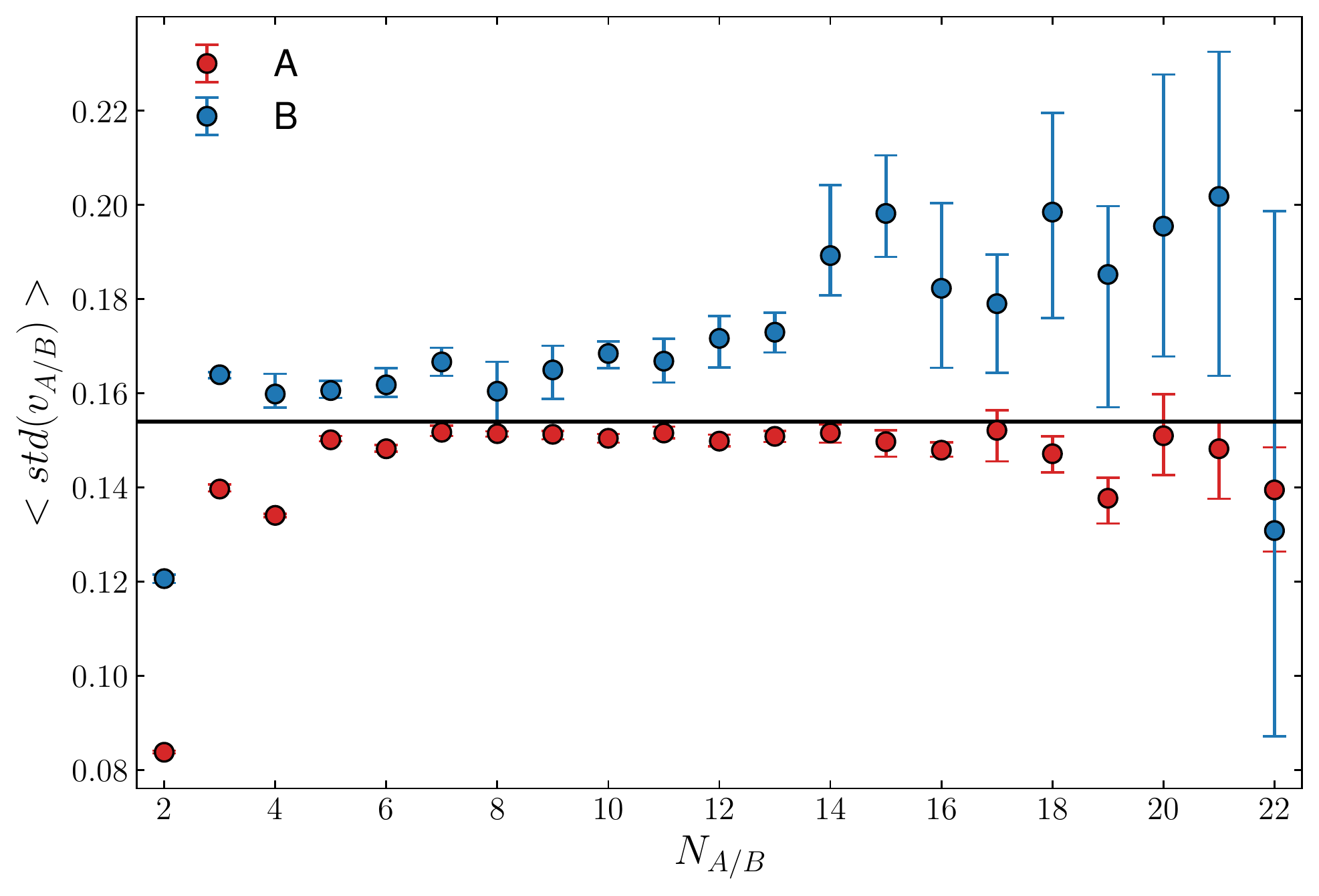}
  \caption{ 
            Average fluctuations of the local pedestrian 
            velocity in dependence of $N_{\rm A}$ and $N_{\rm B}$,
            calculated from the experimental data.
            The plot shows the average standard deviation from the (local) average
            velocity, computed on frames featuring the same number of pedestrians
            respectively in path A and path B.
            The black line represents the global average value of $\sigma = 0.154$.
          }\label{fig:avgstd}
\end{figure}

\paragraph*{Estimation of fluctuations in the velocity fundamental diagram}
In Eq.3 in the main text we define an additive noise term $\epsilon$ for
the fundamental velocity diagram, with $\epsilon$ drawn from a
Gaussian distribution with zero mean and variance $\sigma = 0.15$. 
We have estimated $\sigma$ from the the average fluctuations of 
the local pedestrian velocity in dependence of $N_{\rm A}$
and $N_{\rm B}$, as reported in Fig.~\ref{fig:avgstd}.

In Fig.~\ref{fig:avgstd}, we show the average fluctuations of 
the local pedestrian velocity in dependence of $N_{\rm A}$
and $N_{\rm B}$.
The plot shows the average standard deviation from the (local) average
velocity, computed on frames featuring the same number of pedestrians
respectively in path A and path B. We observe that, within error bars, 
$\sigma$ is constant and independent of $N_{\rm A}$ and $N_{\rm B}$.
%

%=================================================================
%=================================================================

%=================================================================
\section*{Comparison with a different routing policy}
%=================================================================

%=================================================================
\begin{figure}
  \centering
  \includegraphics[width=.49\textwidth]{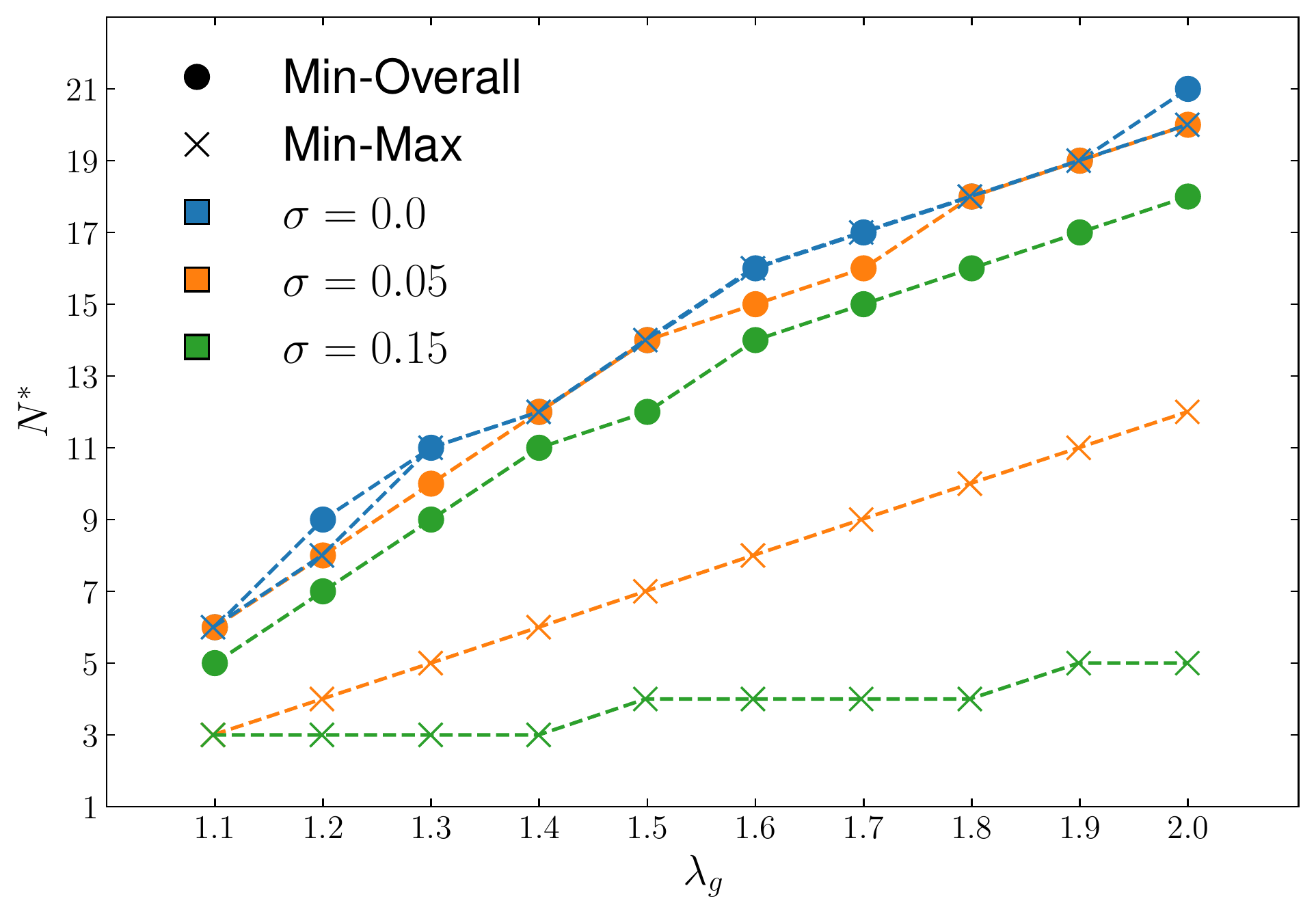}
  \caption{ 
            Comparison of two routing policies, the min-overall policy
            presented in the main text, and the min-max policy included
            in the supplementary material.
            The plot shows the threshold value $N^*$ calculated from simulations
            using different values of $\lambda_g$. 
            We highlighting the effect of introducing fluctuations ($\sigma$) 
            on pedestrians walking speed.
            While the two policies provide very similar results in the deterministic 
            case ($\sigma = 0$), we observe that strong differences arise when including 
            fluctuations.
            In particular, for values of $\sigma$ comparable with those observed in the 
            experimental data ($\sigma = 0.15$), the min-max policy favors the use of path B 
            at a much earlier stage with respect to what observed in the experiment.
            (cf. Fig.~7 in the main text).
          }\label{fig:nstar}
\end{figure}
%=================================================================

%=================================================================
\begin{figure}
  \centering
  \includegraphics[width=.49\textwidth]{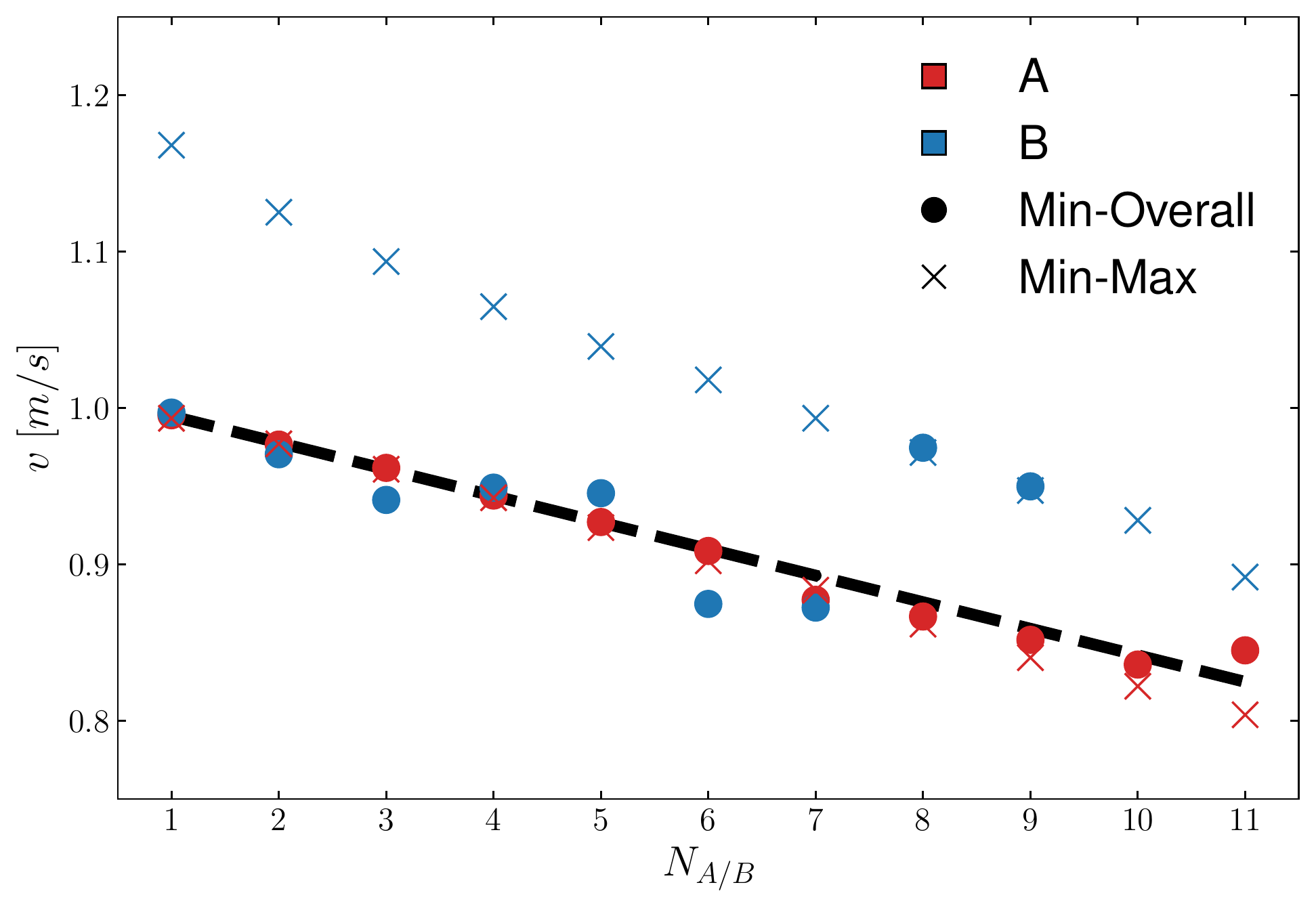}
  \caption{ 
          Local velocity diagram calculated from two simulations
          with $\sigma = 0.1$, $\lambda_g = 1.7$ for the min-max policy,
          and $\lambda_g = 1.3$ for the min-overall policy.
          The black dotted line shows the linear fit of the experimental data,
          which is used in simulations to determine the velocity of pedestrians.
          The plot shows that the min-max policy introduces a systematic 
          selection mechanism, which leads to placing fast walkers in 
          path B, a feature which does not emerge from the experimental data.
          }\label{fig:local-vel-fd}
\end{figure}
%=================================================================

In the main text we have presented numerical results making use 
of a variational principle implementing a policy in which 
pedestrians arrange themselves in order to minimize the 
overall traversal time. This policy is de facto equivalent
to imposing the minimization of the average traversal time.

We here consider an alternative policy, in which pedestrians
perform routing choices by minimizing the worst case scenario,
i.e. the traveling time of the person that takes the longest
to reach destination. The discomfort functional in this case reads as 
\begin{equation}\label{eq:minmax}
  \mathcal{L} = \max_i \tau^{(i)}_{J_i}.
\end{equation}
Even in this case, when neglecting stochastic terms the model
reduces to a Hughes-like form~\cite{hughes2003flow},
with the analytic solution $N_{\rm A}=N_{\rm A}(N)$ defined by the relation 
\begin{equation}\label{eq:deterministic-manifold}
              \frac{L_{\rm A}}{v_{\rm A}(N_{\rm A})} 
  =           \frac{L_{\rm B}}{v_{\rm B}(N_{\rm B})} 
  = \lambda_g \frac{L_{\rm A}}{v_{\rm B}(N_{\rm B})} \quad ,
\end{equation} 
from which it directly follows
\begin{equation}
  \lambda_g =  \frac{v_{\rm B}(N_{\rm B})}{v_{\rm A}(N_{\rm A})} \quad .
\end{equation}
We remark that $\lambda_g$ is equal to the 
velocity ratio between path B and A, 
at variance with the minimum-overall policy reported in the main text
where $\lambda_g$ was instead put in relation with the square
of the velocity ratio.
We take this aspect into account in the comparison of the two 
different policies. In Fig.~\ref{fig:nstar} we plot 
the results of the policy minimizing the worst case scenario
(Eq.~\ref{eq:minmax}, ``min-max'' henceforth)
in correspondence of the
square of value of $\lambda_g$ used in simulations,
in order to make it directly comparable with the min-overall policy.

In Fig.~\ref{fig:nstar} we compute the threshold value $N^*$
for different values of $\lambda_g$, comparing the two
different policies.
When considering the deterministic case 
(respectively Eq.~\ref{eq:deterministic-manifold} and Eq.16 in the main text)
we observe that the two policies provide very similar results.
However, strong differences arise when including 
fluctuations in the local velocity diagram.
When pedestrians with different walking speed are present, 
the min-max policy favors the use of path B at a much 
earlier stage with respect to the min-overall policy;
crucially, the latter provides a more accurate description
of the experimental data, since with $\lambda_g = 1.33$ and 
$\sigma = 0.15$ we correctly reproduce the transition at $N^{*}$
(see again Fig.~7 in the main text).
Reproducing these results with the min-max policy, by accounting 
for the fluctuations observed in the velocity diagram,
would require an (artificially) larger value of $\lambda_g$.

In Fig.~\ref{fig:local-vel-fd} we present a second evidence 
in support to the fact that the min-overall policy provides a 
more accurate description of the experimental data.
In the figure we show a sort of self-consistency check,
by calculating the local velocity diagram from two simulations
with $\sigma = 0.1$, $\lambda_g = 1.7$ for the min-max policy,
and $\lambda_g = 1.3$ for the min-overall policy.
The black dotted line represents the linear fit of the experimental data,
used in simulations to determine the velocity of pedestrians.
The plot shows that the min-max policy introduces a systematic 
selection mechanism, which leads to placing fast walkers in 
path B, a feature which does not emerge from the experimental data
(cf. Fig.~4a in the main text).

%=================================================================
\section*{Relationship between pedestrian count and density}
%=================================================================

In this work we have considered a time-independent modeling approach, 
which has allowed us to neglect complex time correlations 
whose comprehension would have required much more statistics.
Within this framework we have found convenient to take into 
consideration for our analysis the pedestrian count $N$.
As already stated in the main text, this quantity can be put 
in relationship with the density $\rho$ via 
\begin{equation}\label{eq:density-def}
  \rho = \frac{N}{A} ,
\end{equation}
with $A$ the measurement area.

The overall area covered by our depth cameras (Fig.~2) 
consists of approximately $A \approx 28~\rm{m^2}$.
However, using such value for $A$ would lead to an
underestimation of the density since the measured trajectories 
do not uniformly distribute on the measurement area because of
the geometry and typical flow conditions.
%
%=================================================================
\begin{figure}
  \centering
  \includegraphics[width=.49\textwidth]{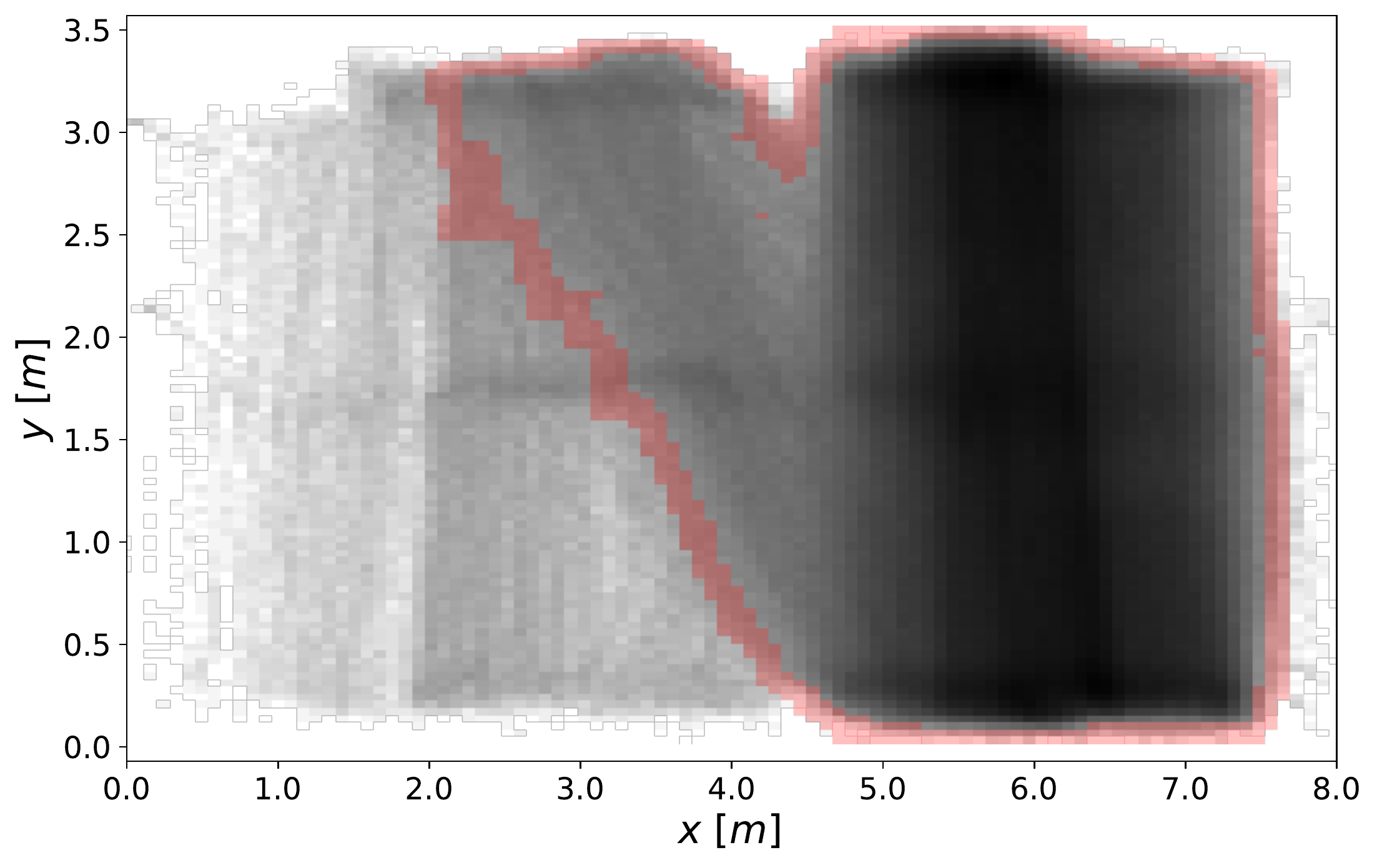}
  \caption{ 
            Boundary (red line) of the reference area ($A_{\rm ref}$) used in 
            the estimation of the pedestrian density ($\rho$).
            The probability distribution function of the pedestrian positions,
            used in the definition of $A_{\rm ref}$,
            is shown in logarithmic scale (grayscale colormap).
          }\label{fig:area}
\end{figure}
%=================================================================
%
Therefore, we compute the density with respect to an 
effective area $A_{\rm ref}$, shaped after the effective floor usage.

In order to calculate $A_{\rm ref}$ we have applied a threshold 
to the probability distribution function of the pedestrian positions 
(2d-histogram in Fig.~1c).
This corresponds to the region bounded by the red line in Fig.~\ref{fig:area},
which represents a domain contributing the $99 \%$ of the occupancy probability.
This yields $A_{\rm ref} = 14.67~\rm{m^2}$.

\end{document}